\documentclass[11pt,english]{article}
\usepackage[top=1in, bottom=1.25in, left=1.25in, right=1.25in]{geometry}
\usepackage{natbib}
\usepackage{amsmath,amssymb,amsfonts,amsthm,mathtools}
\usepackage{bm}
\usepackage{xcolor}
\usepackage{booktabs}
\usepackage{hyperref}
\usepackage[normalem]{ulem} \usepackage{cancel} \usepackage{multirow}
\usepackage{tikz}

\graphicspath{{figures/}{figures/pairwise-vis/}{figures/second-revision/}}

\makeatletter{}
\newtheorem{proposition}{Proposition}

\def\reals{\mathbb{R}}

\def\Blambda{\bm \lambda}
\def\Bmu{\bm \mu}

\renewcommand{\Pr}{\mathbb{P}} 

\definecolor{note}{HTML}{9F1919}
\newcommand{\R}{\mathbb{R}} \newcommand{\Z}{\mathbb{Z}} \newcommand{\given}{\,|\,} 
\newcommand{\nvar}{{d}} \newcommand{\ninst}{{n}} \newcommand{\x}{x}
\newcommand{\xvec}{\bm{\x}}
\newcommand{\xmat}{X}
\newcommand{\nodep}{\theta}
\newcommand{\nodepvec}{\bm{\theta}}
\newcommand{\edgep}{\phi}
\newcommand{\edgepvec}{\bm{\phi}}
\newcommand{\edgepmat}{\Phi}
\newcommand{\domain}{\mathcal{D}}

\newcommand{\cumnormal}{H} 
\newcommand{\condfunc}{\psi} \newcommand{\Cset}{\mathcal{C}} \newcommand{\C}{C} \newcommand{\graph}{\mathcal{G}}
\newcommand{\Eset}{\mathcal{E}} \newcommand{\Vset}{\mathcal{V}} 
\newcommand{\pgmp}{\eta}
\newcommand{\pgmpvec}{\bm{\eta}}

\newcommand{\natp}{\eta}
\newcommand{\Uset}{\mathcal{V}}
\newcommand{\usca}{v}
\newcommand{\uvec}{\mathbf{\usca}}
\newcommand{\radx}{z}

\def\UniP{\Pr_{\text{Poiss}}}
\def\BiP{\Pr_{\text{BiPoi}}}
\def\MulP{\Pr_{\text{MulPoi}}}
\def\MixP{\Pr_{\text{MixedPoi}}}
\def\PLog{\Pr_{\text{PoiLogN}}}
\def\PGM{\Pr_{\text{PGM}}}
\def\TPGM{\Pr_{\text{TPGM}}}
\def\QPGM{\Pr_{\text{QPGM}}}
\def\SPGM{\Pr_{\text{SPGM}}}
\def\FLPGM{\Pr_{\text{FLPGM}}}
\def\SQR{\Pr_{\text{SQR}}}

\def\Var{\text{Var}}

\newlength{\widebarargwidth}
\newlength{\widebarargheight}
\newlength{\widebarargdepth}

\newcommand{\add}[1]{#1}
\newcommand{\removesec}[1]{} \newcommand{\remove}[1]{}

\title{A Review of Multivariate Distributions for Count Data Derived from the Poisson Distribution}
\author{%
David I. Inouye\thanks{The University of Texas at Austin, \url{dinouye@cs.utexas.edu}}, Eunho Yang\thanks{Korea Advanced Institute of Science and Technology, \url{eunhoy@kaist.ac.kr}}, Genevera I. Allen\thanks{Rice University \& Baylor College of Medicine, \url{gallen@rice.edu}}, Pradeep Ravikumar\thanks{Carnegie Mellon University, \url{pradeepr@cs.cmu.edu}}
}
\begin{document}
\maketitle
%
%

\begin{abstract}
The Poisson distribution has been widely studied and used 
for modeling univariate count-valued data.  Multivariate
generalizations of the Poisson distribution that permit dependencies, however, have been far less popular.
Yet, real-world high-dimensional count-valued data found in
word counts, genomics, and crime statistics, for example, exhibit rich
dependencies, and motivate the need for multivariate distributions that
can appropriately model this data. 
We review multivariate distributions derived from the univariate Poisson, categorizing these models into three main classes: 1)
where the marginal distributions are Poisson, 2) where the joint distribution is a
mixture of \remove{Poissons,} \add{independent multivariate Poisson distributions,} and 3) where the node-conditional distributions are
derived from the Poisson.  
We discuss the development of multiple instances of these
classes \add{and compare the models in terms of interpretability and theory.}  Then, we \remove{extensively} \add{empirically} compare multiple models from each class
on \remove{five} \add{three} real-world datasets \add{that have varying data characteristics from different domains, namely } \remove{from} traffic accident data, \remove{crime
statistics,} biological next generation sequencing data, and text \add{data.} \remove{corpora.}  These empirical experiments develop intuition about the comparative advantages and disadvantages of each class of multivariate distribution that was derived from the Poisson. Finally, we suggest new research directions as explored in the subsequent discussion section.

\end{abstract}

\section{Introduction}
Multivariate count-valued data has become increasingly prevalent in
modern big data settings.  Variables in such data are rarely
independent and instead exhibit complex positive and negative
dependencies. We highlight three examples of
multivariate count-valued data that exhibit rich dependencies: text analysis, genomics, and crime statistics.
In text analysis, a standard way to represent documents is to merely
count the number of occurrences for each word in the vocabulary and
create a word-count vector for each document. This representation is
often known as the bag-of-words representation, in which the word order and
syntax are ignored.  The vocabulary size---i.e. \remove{the dimension of the
data---} \add{the number of variables in the data---}is usually much greater than 1000 unique words, and thus a
high-dimensional multivariate distribution is required.  Also, words
are clearly not independent.  For example, if the word ``Poisson''
appears in a document, then the word ``probability'' is \emph{more
  likely} to also appear signifying a positive dependency.  Similarly,
if the word ``art'' appears, then the word ``probability'' is less
likely to also appear signifying a negative dependency.  In genomics,
RNA-sequencing technologies are used to measure gene and isoform
expression levels. These technologies yield counts of reads mapped
back to DNA locations, that even after normalization, yield
non-negative data that is highly skewed with many exact zeros.  This
genomics data is both high-dimensional, with the number of genes measuring
in the tens-of-thousands, and strongly dependent, as genes work
together in pathways and complex systems to produce particular
phenotypes.  In crime analysis, counts of crimes in different counties are clearly multidimensional, with dependencies between crime counts.  For example, the counts of crime in adjacent counties are likely to be correlated with one another, indicating a positive dependency.  While positive dependencies are probably more prevalent in crime statistics, negative dependencies might be very interesting.  For example, a negative dependency between adjacent counties may suggest that a criminal gang has moved from one county to the other.

These examples motivate the need for a high-dimensional count-valued
distribution that permits rich dependencies between variables. In general,
a good class of probabilistic models is a fundamental building block for many tasks 
in data analysis. Estimating such models from data could help answer exploratory questions such as: Which genomic
pathways are altered in a disease e.g. by analyzing genomic
networks?  Or, which county seems to have the strongest
effect, with respect to crime, on other counties?  A probabilistic model could also 
be used in Bayesian classification to determine questions such as:
Does this Twitter post display positive or negative sentiment about a
particular product (fitting one model on positive posts and one model
on negative posts)?  

\begin{figure}[!ht]
\centering
\includegraphics[width=\textwidth]{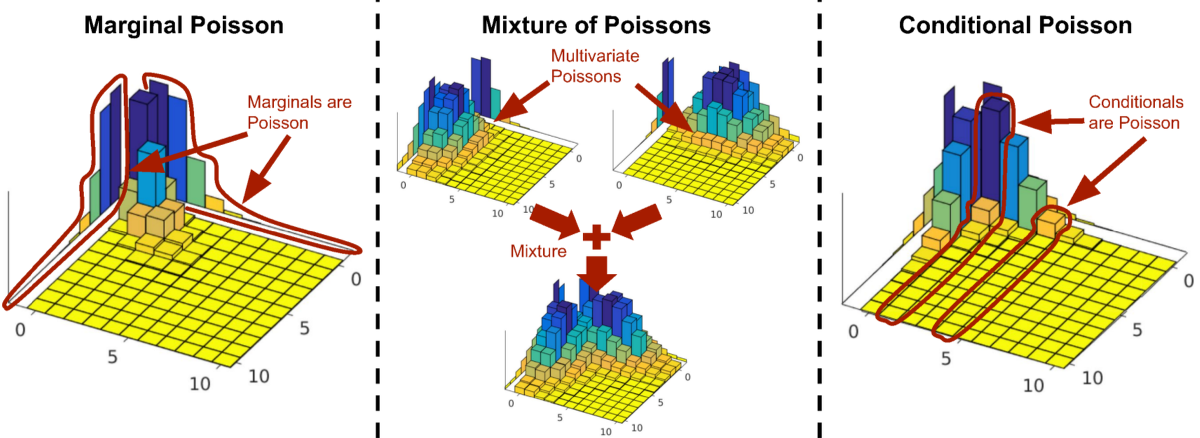}
\caption{(Left) The first class of Poisson generalizations is based on the assumption that the univariate marginals are derived from the Poisson.  (Middle) The second class is based on the idea of mixing independent multivariate Poissons into a joint multivariate distribution. (Right) The third class is based on the assumption that the univariate conditional distributions are derived from the Poisson. }
\label{FigMainClasses}
\end{figure}

The classical model for a count-valued random variable is the
univariate Poisson distribution, whose probability mass function for $\x \in
\{0,1,2,\dots\}$ is:
\add{
\begin{align}\label{EqnUnivPoi}
\UniP( \x \given \lambda ) = \lambda^\x \exp (-\lambda) / \x! \, ,
\end{align}}where $\lambda$ is the standard mean parameter for the Poisson
distribution.  A trivial extension of this to a multivariate distribution would 
be to assume independence between variables, and take the product of node-wise univariate
Poisson distributions, but such a model would
be ill-suited for many examples of multivariate count-valued data that require rich dependence structures.
We review multivariate probability models that are derived from the univariate Poisson distribution and permit non-trivial dependencies between variables. We categorize these models into three main classes based on their primary modeling assumption.  The
first class assumes that the univariate marginal distributions are derived from the Poisson.
The second class is derived as a mixture of independent multivariate
Poisson distributions.  The third class assumes that the univariate
conditional distributions are derived from the Poisson distribution---this last class of models can also be
studied in the context of probabilistic graphical models. An
illustration of each of these three main model classes can be seen in
Fig.~\ref{FigMainClasses}.  While these models might have been classified by primary application area or performance on a particular task, a classification based on modeling assumptions helps emphasize the core abstractions for each model class.  In addition, this categorization may help practitioners from different disciplines learn from the models that have worked well in different areas.  We discuss multiple instances of these
classes in the later sections and highlight the strengths and
weaknesses of each class. \add{We then provide a short discussion on the differences between classes in terms of interpretability and theory.}  Using two different empirical
  measures, we \remove{extensively} \add{empirically} compare multiple models from each class on \remove{five} \add{three} real-world datasets \add{that have varying data characteristics from different domains, namely } \remove{from} traffic accident data, \remove{crime statistics,} biological next generation sequencing data, and text \add{data.} \remove{corpora.}  These experiments develop intuition about the comparative advantages and disadvantages of the models and suggest new research directions as explored in the subsequent discussion section.


\vspace{1em}
\subsection{Notation}
$\R$ denotes the set of real numbers, $\R_+$ denotes the \emph{nonnegative} real numbers, and $\R_{++}$ denotes the \emph{positive} real numbers.  Similarly, $\Z$ denotes the set of integers.  Matrices are denoted as capital letters (e.g. $\xmat, \Phi$), vectors are denoted as boldface lowercase letters (e.g. $\xvec, \bm{\phi}$) and scalar values are non-bold lowercase letters (e.g. $\x, \phi$).

\section{Marginal Poisson Generalizations}\label{SecMarginalPoisson}
The models in this section generalize the univariate Poisson to a multivariate distribution with the property that the marginal distributions of each variable are Poisson.  This is analogous to the marginal property of the multivariate Gaussian distribution, since the marginal distributions of a multivariate Gaussian are univariate Gaussian, and thus seems like a natural constraint when extending the univariate Poisson to the multivariate case.  Several historical attempts at achieving this marginal property have incidentally developed the same class of models, with different derivations~ \citep{m1925applications,campbell1934poisson,wicksell1916some,teicher1954multivariate}.  This marginal Poisson property can also be achieved via the more general framework of copulas~\citep{xue2000multivariate, nikoloulopoulos2009modeling,Nikoloulopoulos2013}.

\subsection{Multivariate Poisson Distribution}\label{SecMultPoi}
\makeatletter{}The formulation of the multivariate Poisson\footnote{The label
``multivariate Poisson'' was introduced in the statistics community to
refer to the particular model introduced in this section but other
generalizations could also be considered multivariate Poisson
distributions.} distribution goes back to \citet{m1925applications}
where authors use differential equations to derive the bivariate
Poisson process. An equivalent but more readable interpretation to
arrive at the bivariate Poisson distribution would be to use the
summation of independent Poisson variables, as
follows \citep{campbell1934poisson}: Let \remove{$\x'_1$, $\x'_2$ and $z$} \add{$y_1$, $y_2$ and $z$} be
univariate Poisson variables with parameters $\lambda_1$, $\lambda_2$
and $\lambda_0$ respectively. Then by setting \remove{$\x_1 = \x'_1 + z$ and
$\x_2 = \x'_2 + z$,} \add{$\x_1 = y_1 + z$ and
$\x_2 = y_2 + z$,} $(\x_1, \x_2)$ follows the bivariate Poisson
distribution, and its joint probability mass is defined as: 
\begin{align}\label{EqnBiPoi}
	& \BiP (\x_1, \x_2 \given \lambda_1, \lambda_2, \lambda_0) \nonumber\\
	=  \ & \exp(-\lambda_1-\lambda_2-\lambda_0) \frac{\lambda_1^{\x_1}}{\x_1!}\frac{\lambda_2^{\x_2}}{_2!} \sum_{z=0}^{\min(\x_1,\x_2)} {{\x_1}\choose{z}} {{\x_2}\choose{z}} z! \left(\frac{\lambda_0}{\lambda_1 \lambda_2}\right)^{z} .
\end{align}

Since the sum of independent Poissons is also Poisson (whose parameter
is the sum of those of two components), the marginal distribution of
$\x_1$ (similarly $\x_2$) is still a Poisson with the rate of
$\lambda_1 + \lambda_0$.  It can be easily seen that the covariance of
$\x_1$ and $\x_2$ is $\lambda_0$ and as a result the correlation
coefficient is somewhere between $0$ and
$\min\{\frac{\sqrt{\lambda_1+\lambda_0}}{\sqrt{\lambda_2+\lambda_0}}, \frac{\sqrt{\lambda_2+\lambda_0}}{\sqrt{\lambda_1+\lambda_0}}\}$ \citep{holgate1964estimation}.
Independently, \citet{wicksell1916some} derived the bivariate Poisson
as the limit of a bivariate binomial
distribution. \citet{campbell1934poisson} show that the models
in \citet{m1925applications} and \citet{wicksell1916some} can
identically be derived from the sums of 3 independent Poisson
variables. 

This approach to directly extend the Poisson distribution can be
generalized further to handle the multivariate case $\xvec \in \Z_+^\nvar$, in which each variable $\x_i$ is the sum of individual Poisson \remove{$\x'_i$} \add{$y_i$} and the common Poisson $\x_0$ as before. The joint probability for a Multivariate Poisson is developed in \citet{teicher1954multivariate} and further considered by other works \citep{dwass1957infinitely,srivastava1970characterization,wang1974characterizations,kawamura1979structure}:
\begin{align}\label{EqnMulPoi}
	\MulP (\xvec ; \Blambda) = \exp \Big(-\sum_{i=0}^\nvar \lambda_i \Big) \Big(\prod_{i=1}^\nvar \frac{\lambda_i^{\x_i}}{\x_i!}\Big) \sum_{z=0}^{\min_i \x_i} \bigg(\prod_{i=1}^\nvar  {{\x_i}\choose{z}}\bigg) z! \left(\frac{\lambda_0}{ \prod_{i=1}^\nvar \lambda_i} \right)^{z} .
\end{align}
Several have shown that this formulation of the multivariate Poisson can also be derived as a limiting distribution of a multivariate binomial distribution when the success probabilities are small and the number of trials is large~\citep{krishnamoorthy1951multivariate,krummenauer1998limit,johnson1997discrete}. 
As in the bivariate case, the marginal distribution of $\x_i$ is
Poisson with parameter $\lambda_i + \lambda_0$. Since $\lambda_0$
controls the covariance between \emph{all} variables, an extremely
limited set of correlations between variables is
permitted.

\citet{mahamunulu1967note} first proposed a more general extension of
the multivariate Poisson distribution that permits a full covariance
structure.  This distribution has been studied further by
many \citep{loukas1983computer,kano1991recurrence,johnson1997discrete,karlis2003algorithm,tsiamyrtzis2004strategies}.
While the form of this general multivariate Poisson 
distribution is too complicated to spell out for $\nvar>3$, its
distribution can be specified by a multivariate reduction scheme.
Specifically, let $y_i$ for $i = 1, \ldots, (2^{\nvar}-1)$ be
independently Poisson distributed with parameter $\lambda_i$.  Now,
define $\mathbf{A} = [ A_1 , A_2, \ldots A_{\nvar} ]$ where $A_i$ is a
$d \times \binom{\nvar}{i}$ matrix consisting of ones and zeros where each
column of $A_i$ has exactly $i$ ones with no duplicate columns.
Hence, $A_1$ is the $\nvar \times \nvar$ identity matrix and
$A_{\nvar}$ is a column 
vector of all ones.  Then,
$\xvec = \mathbf{A} \bm{y}$ is a $\nvar$-dimensional multivariate Poisson
distributed random vector with a full covariance structure.  Note that
the simpler multivariate Poisson distribution with constant covariance
in Eq.~\ref{EqnMulPoi} is a special case of this general form where
$\mathbf{A} = [A_1, A_{\nvar}]$.

The multivariate Poisson distribution has not been widely used for 
real data applications.  This is likely due to two major limitations
of this distribution.  First, the multivariate Poisson
distribution only permits \emph{positive} dependencies; this can
easily be seen as the distribution arises as the sum of independent
Poisson random variables and hence covariances are governed by the
positive rate parameters $\lambda_i$.  The assumption of positive
dependencies is likely unrealistic for most real count-valued data
examples.  
Second, computation of probabilities and inference of parameters is
especially cumbersome for the multivariate Poisson distribution; these
are only computationally tractable for small $\nvar$ and hence not
readily applicable in high-dimensional
settings.  \citet{kano1991recurrence}  
proposed multivariate recursion schemes for computing probabilities,
but these schemes are only stable and computationally feasible for
small $\nvar$, thus complicating likelihood-based inference
procedures.  \citet{karlis2003algorithm}
more recently proposed a latent variable based EM algorithm for
parameter inference of the 
general multivariate Poisson distribution.  This approach treats every
pairwise interaction as a latent variable and conducts inference over
both the observed and hidden parameters.  While this method is more
tractable than recursion schemes, it still requires inference over
$\binom{d}{2}$ latent variables and is hence not feasible in
high-dimensional settings.  Overall,
the multivariate Poisson distribution introduced above is appealing in
that its marginal distributions are Poisson; yet, there are many
modeling 
drawbacks including severe restriction on the types of dependencies
permitted (e.g. only positive relationships), a complicated and 
intractable form in high-dimensions, and challenging inference procedures.

\subsection{Copula Approaches}
A much more general way to construct valid multivariate Poisson distributions with
Poisson marginals is \remove{via \emph{copulas}.} \add{ by pairing a \emph{copula} distribution with Poisson marginal distributions.}
For continuous \remove{marginals} \add{multivariate distributions}, the use of copula \remove{models} \add{distributions}
is founded on the celebrated Sklar's theorem: any continuous joint
distribution can be decomposed into a copula and the marginal
distributions, and conversely, any combination of a copula and
marginal distributions gives a valid continuous joint
distribution \citep{Sklar1959}.  \add{The key advantage of such models for
continuous distributions is that copulas fully specify
the dependence structure hence separating the modeling of marginal
distributions from the modeling of dependencies.} 
\remove{At their core, copula models decouple the modeling of the marginal distributions from the the dependency structure. As will be described in more detail in the following paragraphs, each copula model must specify a copula distribution over the unit hypercube with uniform marginal distributions for each variable (see Fig.~\ref{fig:copula-example} for an example).}
While copula \add{ distributions paired with continuous marginal distributions} enjoy wide popularity \remove{for continuous distributions} (see for example \citep{Cherubini2004} in finance applications), \remove{they} \add{copula models paired with discrete marginal distributions, such as the Poisson,} are more challenging \remove{to
work with for discrete distributions, such as the Poisson,} both for theoretical \remove{reasons} and computational reasons \citep{genest2007primer,Nikoloulopoulos2013a,Nikoloulopoulos2016}.
However, several simplifications and recent advances have attempted to overcome these challenges \citep{Ruschendorf2013,Nikoloulopoulos2013a,Nikoloulopoulos2016}.

\subsubsection{Copula Definition and Examples}
A copula is defined \remove{as} \add{by} a joint cumulative distribution function (CDF),
$C(\bm{u})\colon [0,1]^{\nvar} \to [0,1]$ with uniform marginal
distributions. As a concrete example, the Gaussian
copula (see \add{left subfigure of} Fig.~\ref{fig:copula-example} for an example) is derived from the multivariate normal distribution and is one of the most popular multivariate copulas because of its flexibility in the multidimensional case; the Gaussian copula is defined simply as:
\begin{align*}
\C_R^{\text{Gauss}}( u_1, u_2, \cdots, u_{\nvar}) = \cumnormal_{R} \left( \cumnormal^{-1}( u_1), \cdots,
\cumnormal^{-1}( u_\nvar) \right),
\end{align*}
where $\cumnormal^{-1}(\cdot)$ denotes the standard normal inverse cumulative
distribution function, and $\cumnormal_{R}(\cdot)$ denotes the joint cumulative distribution function of a $\mathcal{N}( 0, R)$ random vector, where $R$ is a correlation matrix.  A similar multivariate copula can be derived from the multivariate Student's $t$ distribution if extreme values are important to model \citep{Demarta2005}.

The Archimedean copulas are another family of copulas which have a \emph{single} parameter that defines the global dependence between all variables \citep{Trivedi2005}.  One property of Archimedean copulas is that they admit an explicit form unlike the Gaussian copula.  Unfortunately, the Archimedean copulas do not directly allow for a rich dependence structure like the Gaussian because they only have one dependence parameter rather than a parameter for each pair of variables.  

\add{Pair copula constructions (PCCs) \citep{aas2009pair} for copulas, or vine copulas, allow combinations of different bivariate copulas to form a joint multivariate copula.  PCCs define multivariate copulas that have an expressive dependency structure like the Gaussian copula but may also model asymmetric or tail dependencies available in Archimedean and $t$ copulas.}
\remove{An altogether different, but related, way of constructing a valid multivariate
distribution is via pair copula constructions (PCCs) \citep{aas2009pair}.}
Pair copulas only use
univariate CDFs, conditional CDFs, and bivariate copulas to construct
a multivariate \remove{density} \add{copula distribution} and hence
can use combinations of the Archimedean copulas described
previously. \add{The multivariate distributions can be factorized in a
variety of ways using bivariate copulas to flexibly model
dependencies.}\remove{ When $d=3$, the
following is an example of a PCC:}\removesec{\begin{align*}
f(x_1, x_2, x_3) &= f_{3|12}(x_3 |
x_1, x_2) \\
& \ \ \times \  f_{2|1}( x_1 | x_2)  \hspace{6mm}  \times  \ f_{1}(x_1) \\
&=  C_{\theta} \left( F_{1|2}(x_1 | x_2), F_{3|2} (x_3 | x_2) \right)
C_{\theta}\left( F_2 (x_2), F_3 (x_3) \right) f_3 (x_3)  \\
& \ \ \times \ C_{\theta}\left( F_2(x_2), F_1(x_1) \right)
f_2(x_2) \ \ \times \ f_1(x_1). 
\end{align*}}\remove{where $C_{\theta}()$ is a bivariate copula.}\emph{Vines}, or graphical
tree-like structures, denote the possible factorizations \remove{of a
multivariate densities} that are feasible for PCCs \remove{; in a
vine, each edge 
in the tree structure is parameterized by a bivariate copula}
\citep{bedford2002vines}.

\subsubsection{Copula Models for Discrete Data}
As per Sklar's theorem, any copula distribution can be combined with marginal distribution CDFs $\{F_i(\x_i)\}_{i=1}^{d}$ to create a joint distribution:
\begin{align*}
G( \x_1, \x_2, \cdots,  \x_{\nvar} \given \theta, F_1,\cdots,F_\nvar ) = C_{\theta} \left( u_1=F_{1}( \x_{1}), \cdots, u_\nvar=F_{\nvar}( \x_{\nvar})\right).\,
\end{align*}
If sampling from the given copula is possible, this form admits simple direct sampling from the joint distribution (defined by the CDF $G(\cdot)$) by first sampling from the copula $\bm{u} \sim \text{Copula}(\theta)$ and then transforming $\bm{u}$ to the target \remove{domain}\add{space} using the inverse CDFs of the marginal distributions: $\bm{x} = [ F_1^{-1}(u_1), \cdots, F_\nvar^{-1}(u_\nvar) ]$.

\begin{figure}[!ht]
\centering
\includegraphics[width=\textwidth]{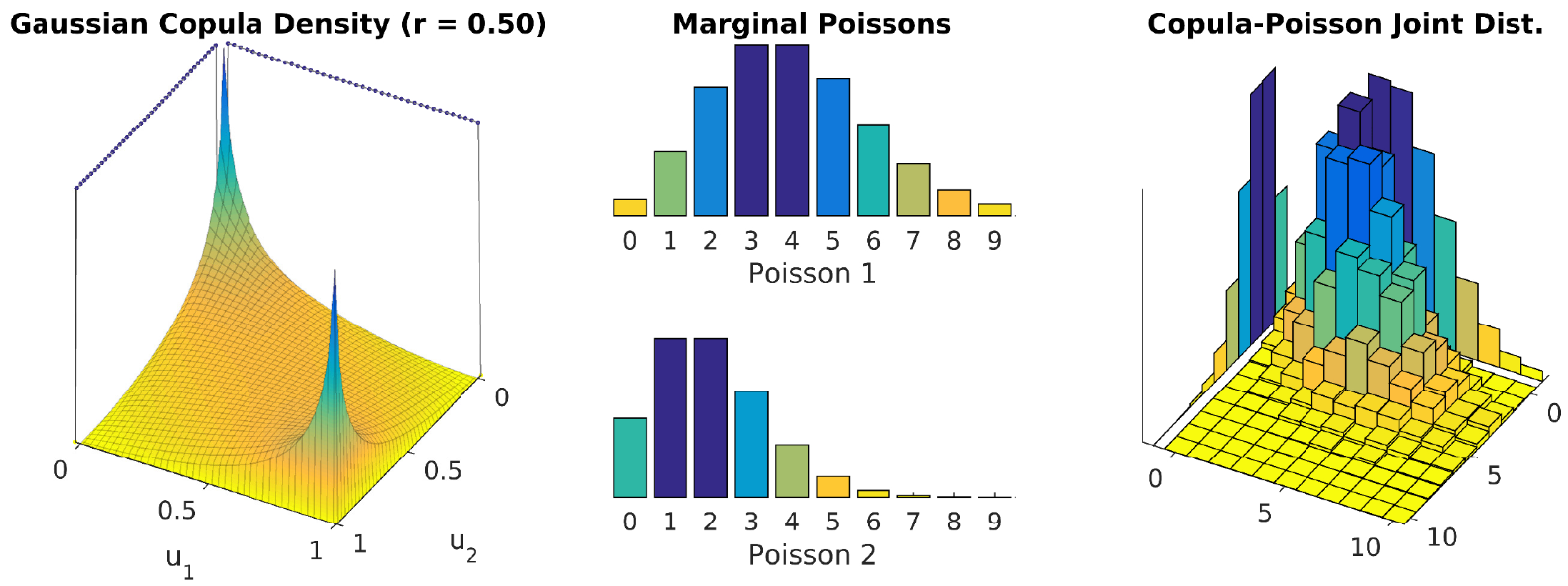}
\caption{\removesec{ \sout{A copula model has two components that together define a joint distribution:}} \add{A copula distribution (left)---which is defined over the unit hypercube and has uniform marginal distributions---,}\removesec{\sout{whose domain is the unit hypercube ($[0,1]^\nvar$) and whose marginal distributions are uniform,}} \add{paired with  univariate Poisson} marginal distributions for each \removesec{\sout{dimension}} \add{variable (middle) defines a valid discrete joint distribution with Poisson marginals (right).}}
\label{fig:copula-example}
\end{figure}

A valid multivariate \add{discrete} joint distribution can be derived \add{by pairing a copula distribution with Poisson marginal distributions.}\remove{using copulas by letting the marginals be discrete.} For example, a valid joint CDF with Poisson marginals is given by
\begin{align*}
G( \x_1, \x_2, \cdots,  \x_{\nvar} \given \theta ) = C_{\theta} \left( F_{1}( \x_{1} \given \lambda_{1}), \cdots, F_{\nvar}( \x_{\nvar}\given \lambda_{\nvar})\right),
\end{align*}
where $F_i( \x_i \given \lambda_i)$ is the Poisson cumulative
distribution function with mean parameter $\lambda_i$, and $\theta$
denotes the copula parameters.  \remove{If we use the Gaussian copula, then we have a Poisson-Gaussian copula construction} \add{If we pair a Gaussian copula with Poisson marginal distributions, we create a valid joint distribution} that has been widely 
used for generating samples \remove{from}\add{of} multivariate count data~\citep{xue2000multivariate,yahav2012generating,cook2010copula}\add{---an example of the Gaussian copula paired with Poisson marginals to form a discrete joint distribution can be seen in Fig.~\ref{fig:copula-example}}.

\citet{Nikoloulopoulos2013} present an excellent survey of \remove{discrete copula models}\add{copulas to be paired with discrete marginals} by defining several desired properties of a copula (quoted from \citep{Nikoloulopoulos2013}):
\begin{enumerate}
\item Wide range of dependence, allowing both positive and negative dependence.
\item Flexible dependence, meaning that the number of bivariate marginals is (approximately) equal to the number of dependence parameters.
\item Computationally feasible cumulative distribution function (CDF) for likelihood estimation.
\item Closure property under marginalization, meaning that lower-order marginals belong to the same parametric family.
\item No joint constraints for the dependence parameters, meaning that the use of covariate functions for the dependence parameters is straightforward.
\end{enumerate}
Each copula model satisfies some of these properties but not all of them.  For example, Gaussian copulas satisfy properties (1), (2) and (4) but not (3) or (5) because the normal CDF is not known in closed form and the positive definiteness constraint on the correlation matrix.  \citet{Nikoloulopoulos2013} recommend Gaussian copulas for general models and vine copulas if modeling dependence in the tails or asymmetry is needed.

\subsubsection{Theoretical Properties of \add{Copulas Derived from Discrete Distributions}\removesec{Discrete Copulas}}
From a theoretical perspective, \remove{discrete copulas exist but
}  \add{a multivariate discrete distribution can be viewed as a
continuous copula distribution paired with discrete marginals but the
derived copula distributions} are not
unique \add{ and hence, are unidentifiable~\citep{genest2007primer}.
Note that this is in contrast to 
continuous multivariate distributions where the derived copulas are uniquely
defined~\citep{sklar1973random}.}
Because of this
non-uniqueness property, \citet{genest2007primer} caution against
performing inference \remove{in discrete copula models} \add{on}
and interpreting
dependencies \add{of copulas derived from discrete
distributions.} \remove{in discrete copula models.} \add{A further
consequence of non-uniqueness is that when
copula distributions are paired with discrete marginal distributions,
the copulas no longer fully specify the dependence structure as with
continuous marginals \citep{genest2007primer}.  In other words, the
dependencies of the joint distribution will depend in part on which
marginal distributions are employed.  In practice, this often means
that the range of dependencies permitted with certain copula and
discrete marginal distribution pairs is much more limited than the
copula distribution would otherwise model.
However, several have suggested that this non-uniqueness property does
not have major practical
ramifications \citep{Nikoloulopoulos2013,Karlis2016}. }
\remove{However, in practice, this non-uniqueness property does not seem to have significant practical ramifications \citep{Nikoloulopoulos2013,Karlis2016}.}

\add{We discuss a few common approaches used for the estimation of continuous copulas with discrete marginals.}

\subsubsection{Continuous Extension for Parameter Estimation}
For estimation of continuous copulas from data, a two-stage procedure called Inference Function for Marginals (IFM) \citep{Joe1996} is commonly used in which the marginal distributions are estimated first and then used to map the data onto the unit hypercube using the CDFs of the inferred marginal distributions.  While this is straightforward for continuous marginals, this procedure is less obvious for discrete marginal distributions when using a continuous copula.  One idea is to use the continuous extension (CE) of integer variables to the continuous domain \citep{Denuit2005} by forming a new ``jitter'' continuous random variable $\tilde{x}$:
\begin{align*}
\tilde{x} = x + (u - 1) \, ,
\end{align*}
where $u$ is a random variable defined on the unit interval.  It is straightforward to see that this new random variable is continuous and $\tilde{x} \leq x$.  An obvious choice for the distribution of $u$ is the uniform distribution.  With this idea, inference can be performed using a surrogate likelihood by randomly projecting each discrete data point into the continuous domain and averaging over the random projections as done in \citep{Heinen2007,Heinen2008}.  \citet{Madsen2009,Madsen2011} use the CE idea as well but generate \remove{new samples} \add{\emph{multiple} jittered samples $\{\tilde{x}^{(1)},\tilde{x}^{(1)},\dots,\tilde{x}^{(m)} \}$ for each original observation $x$} to estimate the discrete likelihood rather than merely \add{generating one jittered sample $\tilde{x}$ for each original observation $x$} \remove{jittering the original data} as in \citep{Heinen2007,Heinen2008}.  \remove{\citet{Nikoloulopoulos2013a} compare to these CE methods and demonstrate that doing discrete maximum likelihood estimation (MLE) can be well approximated by using a different simulation scheme based on estimating multivariate normal rectangular probabilities.}  \citet{Nikoloulopoulos2013a} find that CE-based methods significantly underestimate the correlation structure because the CE jitter transform operates independently for each variable instead of considering the correlation structure between the variables.

\subsubsection{Distributional Transform for Parameter Estimation}
In a somewhat different direction, \citet{Ruschendorf2013} proposed the use of a generalization of the CDF distribution function $F(\cdot)$ for the case with discrete variables, which they term a  \emph{distributional transform} (DT) denoted by $\tilde{F}(\cdot)$:
\begin{align*}
\tilde{F}(x,v) \equiv F(x) + v\Pr(x) = \Pr(X < x) + v\Pr(X = x) \,,
\end{align*}
where $v \sim \text{Uniform}(0,1)$.  Note that in the continuous case, $\Pr(X = x) = 0$ and thus this reduces to the standard CDF for continuous distributions. One way of thinking of this modified CDF is that the random variable $v$ adds a random jump when there are discontinuities in the original CDF.  If the distribution is discrete (or more generally if there are discontinuities in the original CDF), this transformation enables a \add{simple} proof of a theorem akin to Sklar's theorem for discrete distributions \citep{Ruschendorf2013}. \remove{While this does not circumvent the problems of non-uniqueness of the copula, it does provide a theoretical foundation for discrete copulas.}

\citet{Kazianka2010,Kazianka2013} propose using the distributional transform (DT) from \citep{Ruschendorf2013} to develop a simple and intuitive approximation for the likelihood.  Essentially, they simply take the expected jump value of $\mathbb{E}(v) = 0.5$ (where $v \sim \text{Uniform}(0,1)$) and thus transform the discrete data to the continuous domain by the following:
\begin{align*}
u_i \equiv F_i(x_i-1) + 0.5\Pr(x_i) = 0.5( F_i(x_i-1) + F_i(x_i) ) \,,
\end{align*}
which can be seen as simply taking the average of the CDF values at $x_i -1$ and $x_i$.  Then, they use a continuous copula \remove{ model} such as the Gaussian copula.  Note that this is much simpler to compute than the simulated likelihood (SL) method in \citep{Nikoloulopoulos2013a} or the continuous extension (CE) methods in \citep{Heinen2007,Heinen2008,Madsen2009,Madsen2011}, which require averaging over many different random initializations.

\subsubsection{Simulated Likelihood for Parameter Estimation}
Finally, \cite{Nikoloulopoulos2013a} propose a method to directly approximate the maximum likelihood estimate by estimating a discretized Gaussian copula.  Essentially, unlike the CE and DT methods which attempt to transform discrete variables to continuous variables, \add{the MLE for a Gaussian copula with discrete marginal distributions $F_1,F_2,\dots,F_\nvar$ can be formulated as estimating multivariate normal rectangular probabilities:
\begin{align}
\Pr(\xvec \given \bm{\gamma}, R) &= \int^{\phi^{-1}[F_1(x_1 \given \gamma_1)]}_{\phi^{-1}[F_1(x_1 -1 \given \gamma_1)]} \cdots \int^{\phi^{-1}[F_\nvar(x_\nvar \given \gamma_\nvar)]}_{\phi^{-1}[F_1(x_\nvar -1 \given \gamma_\nvar)]} \Phi_R(z_1,\dots,z_\nvar) \text{d}z_1\dots\text{d}z_\nvar \, ,
\end{align}
where $\bm{\gamma}$ are the marginal distribution parameters, $\phi^{-1}(\cdot)$ is the univariate standard normal inverse CDF, and $\Phi_R(\cdots)$ is the multivariate normal density with correlation matrix $R$.
}
\remove{---i.e. the discrete probabilities corresponding to the discrete marginals---}
\add{\citet{Nikoloulopoulos2013a} propose to approximate the multivariate normal rectangular probabilities via} fast simulation algorithms discussed in \citep{Genz2009}.  \add{Because this method directly approximates the MLE via simulated algorithms,} this method is called simulated likelihood (SL). \citet{Nikoloulopoulos2016} compare the DT and SL methods for small sample sizes and find that the DT method tends to overestimate the correlation structure.  However, because of the computational simplicity, \citet{Nikoloulopoulos2016} give some heuristics of when the DT method might work well compared to the more accurate but more computationally expensive SL method.

\add{
\subsubsection{Vine Copulas for Discrete Distributions}
\citet{panagiotelis2012pair} provide conditions under which a multivariate discrete distribution can be decomposed as a vine PCC copula paired with discrete marginals.  In addition, \citet{panagiotelis2012pair} show that likelihood computation for vine PCCs with discrete marginals is quadratic as opposed to exponential as would be the case for general multivariate copulas such as the Gaussian copula with discrete marginals.
}
\remove{Recently, \citet{panagiotelis2012pair} introduced discrete vine PCCs
by generalizing well-studied continuous vine PCCs to the discrete domain.  
These discrete vine copulas offer two key advantages including more flexible
modeling that permits asymmetries and tail dependence and faster
computation of probabilities and parameters \citep{panagiotelis2012pair}.} 
However, computation in truly high-dimensional settings remains a
challenge as $2d (d-1)$ bivariate copula evaluations are required to
calculate the PMF or likelihood of a $d$-variate PCC using the
algorithm proposed by \citet{panagiotelis2012pair}.  These bivariate
copula evaluations, however, can be coupled with some of the previously
discussed computational techniques such as continuous extensions,
distributional transforms, and simulated likelihoods for further
computational improvements.  Finally, while vine PCCs offer a very 
flexible modeling approach, this comes with the added challenge of
selecting the vine construction and bivariate copulas
\citep{czado2013selection}, which has not been well studied for discrete distributions.
Overall, \citet{Nikoloulopoulos2013} recommend using \remove{discrete} vine PCCs
for complex modeling \add{of discrete data} with tail dependencies and asymmetric dependencies.

\subsection{Summary of Marginal Poisson Generalizations}
We have reviewed the historical development of the multivariate Poisson which has Poisson marginals and then reviewed many of the recent developments of using the much more general copula framework to derive Poisson generalizations with Poisson marginals.  The original multivariate Poisson models based on latent Poisson variables are limited to positive dependencies and require computationally expensive algorithms to fit.  However, estimation of \remove{the discrete copula models} \add{copula distributions paired with Poisson marginals}---while theoretically has some caveats---can be performed efficiently in practice.  Simple approximations such as the expectation under the distributional transformation can provide nearly trivial transformations that move the discrete variables to the continuous domain in which all the tools of continuous copulas can be exploited.  More complex transformations such as the simulated likelihood method \citep{Nikoloulopoulos2013a} can be used if the sample size is small or high accuracy is needed.

\section{Poisson Mixture Generalizations}
Instead of directly extending univariate Poissons to the multivariate case, a separate line of work proposes to indirectly extend the Poisson based on the mixture of independent Poissons. Mixture models are often considered to provide more flexibility by allowing the parameter to vary according to a mixing distribution.
One important property of mixture models is that they can model \emph{overdispersion}.  Overdispersion occurs when the variance of the data is larger than the mean of the data---unlike in a Poisson distribution in which the mean and variance are equal.  One way of quantifying dispersion is the dispersion index:
\begin{align}
\delta = \frac{\sigma^2}{\mu} \, .
\label{eqn:dispersion}
\end{align}
If $\delta > 1$, then the distribution is overdispersed whereas if $\delta < 1$, then the distribution is underdispersed.  In real world data as will be seen in the experimental section, overdispersion is more common than underdispersion.  Mixture models also enable dependencies between the variables as will be described in the following paragraphs.

Suppose that we are modeling univariate random variable $\x$ with a density of $f(\x \given \theta)$. Rather than assuming $\theta$ is fixed, we let $\theta$ itself to be a random variable following some \emph{mixing} distribution. More formally, a general \emph{mixture} distribution can be defined as \citep{karlis2005mixed}:
\begin{align}\label{EqnMixture}
	\Pr(\x \given g(\cdot))  = \int_\Theta f(\x \given \theta ) \,  g(\theta) \, d \theta  \, ,
\end{align}
where the parameter $\theta$ is assumed to come from the mixing distribution $g(\theta)$ and $\Theta$ is the domain of $\theta$.

For the Poisson case, let $\Blambda \in \R_{++}^\nvar$ be a $\nvar$-dimensional vector whose $i$-th element $\lambda_i$ is the parameter of the Poisson distribution for $\x_i$. Now, given some mixing distribution $g(\Blambda)$, the family of Poisson mixture distributions is defined as
\begin{align}\label{EqnMixedPoi}
	\MixP (\xvec) = \int_{\R_{++}^d}  g ( \Blambda ) \, \prod_{i=1}^d \UniP( \x_i \given \lambda_i ) \, d \Blambda \, ,
\end{align} 
where the domain of the joint distribution is any count-valued assignment (i.e. $\x_i \in \Z_+, \forall i$).  While the probability density function (Eq.~\ref{EqnMixedPoi}) has the complicated form involving a multidimensional integral (a complex, high-dimensional integral when $\nvar$ is large), the mean and variance are known to be expressed succinctly as
\begin{align}
	\mathbb{E}(\xvec) & = \mathbb{E}(\Blambda) \, , \\
	\Var (\xvec) & =  \mathbb{E}(\Blambda) + \Var (\Blambda) \, . \label{EqnMixedPoiProperty}
\end{align}
Note that Eq.~\ref{EqnMixedPoiProperty} implies that the variance of a mixture is always larger than the variance of a single distribution.  The higher order moments of $\xvec$ are also easily represented by those of $\Blambda$. Besides the moments, other interesting properties (convolutions, identifiability etc.) of Poisson mixture distributions are extensively reviewed and studied in \citet{karlis2005mixed}.

One key benefit of Poisson mixtures is that they permit both positive as well as negative dependencies simply by properly defining $g(\Blambda)$. The intuition behind these dependencies can be more clearly understood when we consider the sample generation process. Suppose that we have the distribution $g(\Blambda)$ \add{in two dimensions (i.e. $d=2$)} with a strong positive dependency between $\lambda_1$ and $\lambda_2$. Then, given a sample $(\lambda_1,\lambda_2)$ from $g(\Blambda)$, $\x_1$ and $\x_2$ are likely to also be positively correlated.

In an early application of the model, 
\citet{arbous1951accident} constrain the Poisson parameters as the different scales of common gamma variable $\lambda$: for $i=1,\hdots,\nvar$, the time interval $t_i$ is given and $\lambda_i$ is set to $t_i \lambda$. Hence, $g(\Blambda)$ is a univariate gamma distribution specified by $\lambda \in \R_{++}$
---which only allows simple dependency structure. \citet{steyn1976multivariate}, as another early attempt, choose the multivariate normal distribution for the mixing distribution $g(\Blambda)$ to provide more flexibility on the correlation structure. However, the normal distribution poses problems because $\lambda$ must reside in $\R_{++}$ while the the normal distribution is defined on $\R$.

One of the most popular choice for $g(\Blambda)$ is the log-normal
distribution thanks to its rich covariance structure and natural positivity constraint\footnote{This is because if $y \in \R \sim\text{Normal}$, then $\exp(y) \in \R_{++} \sim \text{LogNormal}$.}:
\begin{align}
	\mathcal{N}_{\log} ( \Blambda \given \Bmu, \Sigma ) = \frac{1}{\Pi_{i=1}^d \lambda_i \sqrt{(2\pi)^d |\Sigma|}} \exp \Big( -\frac{1}{2} (\log \Blambda - \Bmu)^\top \Sigma^{-1} (\log \Blambda - \Bmu)  \Big) \, .
\end{align}
The log-normal distribution above is parameterized by $\Bmu$ and $\Sigma$, which are the mean and the covariance of $(\log \lambda_1, \log \lambda_2, \cdots, \log \lambda_d)$, respectively. Setting the random variable $\x_i$ to follow the Poisson distribution with parameter $\lambda_i$, we have the multivariate Poisson log-normal distribution \citep{aitchison1989multivariate} from Eq.~\ref{EqnMixedPoi}:
\add{
\begin{align}\label{EqnPoiLogNormal}
	\PLog (\xvec \given \Bmu , \Sigma) = \int_{\reals_{+}^d} \mathcal{N}_{\log} ( \Blambda \given \Bmu, \Sigma ) \, \prod_{i=1}^d \UniP( \x_i \given \lambda_i ) \, d \Blambda \, .
\end{align}}While the joint distribution (Eq.~\ref{EqnPoiLogNormal}) does not have a closed-form expression and hence  as $d$ increases, it becomes computationally cumbersome to work with, its moments are available in closed-form as a special case of Eq.~\ref{EqnMixedPoiProperty}:
\begin{align}
	\alpha_i \equiv \mathbb{E}(\x_i) & = \exp(\mu_i + \frac{1}{2} \sigma_{ii}) \, , \nonumber\\	
	\Var (\x_i) & =  \alpha_i + \alpha_i^2 \big(\exp(\sigma_{ii}) - 1 \big)  \, , \nonumber\\
	{\text{Cov}}(\x_i,\x_j) & = \alpha_i \alpha_j (\exp(\sigma_{ij}) - 1 \big) \, .
	\label{EqnCovMixture}
\end{align}
The correlation and the degree of overdispersion (defined as the 
variance divided by the mean) of the marginal distributions are strictly coupled by $\alpha$ and $\sigma$. Also, the possible Spearman's $\rho$ correlation values for this distribution are limited if the mean value $\alpha_i$ is small.  To briefly explore this phenomena, we simulated a two-dimensional Poisson log-normal model with mean zero and covariance matrix:
\begin{align*}
\Sigma = 2\log(\alpha_i)\begin{bmatrix} 1 & \pm 0.999 \\ \pm 0.999 & 1 \end{bmatrix}\, ,
\end{align*}
which corresponds to a mean value of $\alpha_i$ per Eq.~\ref{EqnCovMixture} and the strongest positive and negative correlation possible between the two variables.  We simulated one million samples from this distribution and found that when fixing $\alpha_i = 2$, the Spearman's $\rho$ values are between -0.53 and 0.58.  When fixing $\alpha_i = 10$, the Spearman's $\rho$ values are between -0.73 and 0.81.  Thus, for small mean values, the log-normal mixture is limited in modeling strong dependencies but for large mean values the log-normal mixture can model stronger dependencies.
Besides the examples provided here, various Poisson mixture models from different mixing distributions are available although limited in the applied statistical literature due to their complexities. See \citet{karlis2005mixed} and the references therein for more examples of Poisson mixtures. \citet{karlis2005mixed} also provide the general properties of mixtures as well as the specific ones of Poisson mixtures such as moments, convolutions, and the posterior.

While this review focuses on modeling multivariate count-valued responses without any extra information, the several extensions of multivariate Poisson log-normal models have been proposed to provide more general correlation structures when covariates are available \citep{Chib2001,ma2008multivariate,park2007multivariate,el2009collision,Aguero-Valverde2009,Zhan2015}. These works formulate the mean parameter of log-normal mixing distribution, $\log \mu_i$, as a linear model on given covariates in the Bayesian framework. 

In order to alleviate the computational burden of using log-normal distributions as an infinite mixing density as above, \citet{karlis2007finite} proposed an EM type estimation for a finite mixture of $k > 1$ Poisson distributions, which still preserves similar properties such as both positive and negative dependencies, as well as closed form moments.  While \citep{karlis2007finite} consider mixing multivariate Poissons with positive dependencies, the simplified form where the component distributions are independent Poisson distributions is much simpler to implement using an expectation-maximization (EM) algorithm. This simple finite mixture distribution can be viewed as a middle ground between a single Poisson and a non-parametric estimation method where a Poisson is located at every training point---i.e. the number of mixtures is equal to the number of training data points ($k=\ninst$).

The gamma distribution is another common mixing distribution for the Poisson because it is the conjugate distribution for the Poisson mean parameter $\lambda$. For the univariate case, if the mixing distribution is gamma, then the resulting univariate distribution is the well-known negative binomial distribution.  The negative binomial distribution can handle overdispersion in count-valued data when the variance is larger than the mean.  Unlike the Poisson log-normal mixture, the univariate gamma-Poisson mixture density---i.e. the negative binomial density---is known in closed form:
\begin{align*}
\Pr( x \given r, p) = \frac{\Gamma(r+x)}{\Gamma(r)\Gamma(x+1)} p^r (1-p)^x \,.
\end{align*}
As $r \to \infty$, the negative binomial distribution approaches the Poisson distribution.  Thus, this can be seen as a generalization of the Poisson distribution.  Note that the variance of this distribution is always larger than the Poisson distribution with the same mean value.

In a similar vein to using the gamma distribution, if instead of putting a prior on the Poisson mean parameter $\lambda$, we reparametrize the Poisson distribution by the log Poisson mean parameter $\theta = \log(\lambda)$, then the log-gamma distribution is conjugate to parameter $\theta$.
\citet{bradley2015computationally} recently leveraged the log-gamma conjugacy to the Poisson log-mean parameter $\theta$ by introducing the Poisson log-gamma hierarchical mixture distribution.  In particular, they discuss the multivariate log-gamma distribution that can have flexible dependency structure similar to the multivariate log-normal distribution and illustrate some modeling advantages over the log-normal mixture model.

\subsection{Summary of Mixture Model Generalizations}
Overall, mixture models are particularly helpful if there is overdispersion in the data---which is often the case for real-world data as seen in the experiments section---while also allowing for variable dependencies to be modeled implicitly through the mixing distribution.  If the data exhibits overdispersion, then the log-normal or log-gamma distributions \citep{bradley2015computationally} give somewhat flexible dependency structures.  The principal caveat with complex mixture of Poisson distributions is computational; exact inference of the parameters is typically computationally difficult due to the presence of latent mixing variables.  However, simpler models such as the finite mixture using simple expectation maximization (EM) may provide good results in practice (see comparison section).

\section{Conditional Poisson Generalizations}
While the multivariate Poisson formulation in Eq.~\ref{EqnMulPoi} as well as the \add{distribution formed by pairing a copula with Poisson marginals} \remove{copula-based generalizations}
assume that univariate \emph{marginal distributions} are derived from the Poisson, a
different line of work generalizes the univariate Poisson by assuming
the univariate \emph{node-conditional distributions} are derived from the Poisson \citep{Besag74,YRAL12,YRAL13a,YRAL15,IRD15,IRD16}.  Like the
assumption of Poisson marginals in previous sections, this conditional
Poisson assumption seems a different yet natural extension of
the univariate Poisson distribution. The multivariate Gaussian can be seen to satisfy such a conditional property since the node-conditional distributions of a multivariate Gaussian are univariate Gaussian. One benefit of these conditional models is that they can be seen as undirected graphical models or
Markov Random Fields, and they have a simple parametric form.  In addition, \remove{learning}\add{estimating} these models
generally reduces to estimating simple node-wise regressions, and some
of these estimators have theoretical guarantees on \remove{learning}\add{estimating} the global graphical model structure even under high-dimensional sampling regimes, where the number of variables ($\nvar$) is potentially even larger than the number of samples ($\ninst$).

\subsection{Background on Exponential Family Distributions}
We briefly describe exponential family distributions and graphical models which form the
basis for the conditional Poisson models. Many commonly used
distributions fall into this family, including Gaussian, Bernoulli,
exponential, gamma, and Poisson, among others.  The exponential family is specified by
a vector of sufficient statistics denoted by $T(\xvec) \equiv [T_1(\xvec),
T_2(\xvec), \cdots, T_m(\xvec)]$, the log base measure $B(\xvec)$ and
the domain of the random variable $\domain$.  With this notation, the
generic exponential family is defined as: 
\begin{align*}
\Pr_{\text{ExpFam}}(\xvec \given \pgmpvec) &= \exp\left( \sum_{i=1}^m \pgmp_i T_i(\xvec) + B(\xvec) - A(\pgmpvec) \right)\\
A(\pgmpvec) &=\log \int_\domain \exp\left( \sum_{i=1}^m \pgmp_i T_i(\xvec) + B(\xvec) \right)\, \mathrm{d}\mu(\xvec) \, ,
\end{align*}
where $\pgmpvec$ are called the \emph{natural or canonical parameters}
of the distribution, $\mu$ is the Lebesgue or counting measure
depending on whether $\domain$ is continuous or discrete respectively, 
and $A(\pgmpvec)$ is called the \emph{log partition function}
or \emph{log normalization constant} because it normalizes the
distribution over the domain $\domain$. Note that
the sufficient statistics $\{T_i(\xvec)\}_{i=1}^{m}$ can be any arbitrary function
of $\xvec$; for example, $T_i(\xvec) = \x_1\x_2$ could be used to model
interaction between $\x_1$ and $\x_2$.  The log partition function
$A(\pgmpvec)$ will be a key quantity when discussing the following models: 
$A(\pgmpvec)$ must be finite for the distribution to be valid, so that the realizable domain of parameters is given by 
$\{\pgmpvec \in \domain\,:\, A(\pgmpvec) < \infty\}$. Thus, for instance, if the realizable domain only allows positive or negative interaction terms,\remove{ for instance,} \add{then the set of allowed dependencies would be severely restricted.} \remove{that would severely restrict the set of allowed dependencies in the model.} 

Let us now consider the exponential family form of the univariate Poisson:
\begin{align}
\UniP( \x \given \lambda ) &= \lambda^\x \exp (-\lambda) / \x!  \notag \\
&=\exp(\log(\lambda^\x) - \log(\x!) -\lambda) \notag \\
&=\exp(\underbrace{\log(\lambda)}_{\pgmp} \underbrace{\x}_{T(\x)} + \underbrace{(-\log(\x!))}_{B(\x)} - \lambda) \, , \quad \text{and therefore}\notag \\
\UniP( \x \given \pgmp ) &=\exp(\pgmp \x -\log(\x!) - \exp(\pgmp) ) \, ,
\label{EqnUnivPoissExpFam}
\end{align}
where $\pgmp \equiv \log(\lambda)$ is the natural parameter of the Poisson, $T(\x) = \x$ is the Poisson sufficient statistic, $-\log(\x!)$ is the Poisson log base measure and $A(\pgmp) = \exp(\pgmp)$ is the Poisson log partition function.  Note that for the general exponential family distribution, the log partition function may not have a closed form.

\subsection{Background on Graphical Models}
The graphical model over $\xvec$ given some graph $\graph$---a set of nodes and edges---is a set of distributions on $\xvec$ that satisfy the Markov independence assumptions with respect to $\graph$~\citep{lauritzen_1996}. In particular, an undirected graphical model gives a compact way to represent \emph{conditional} independence among random variables---the \emph{Markov} properties of the graph.  Conditional independence relaxes the notion of full independence by defining which variables are independent \emph{given} that other variables are fixed or known.
  
More formally, let $\graph = (\Vset,\Eset)$ be an undirected graph over $\nvar$ nodes in $\Vset$ corresponding $\nvar$ random variables in $\xvec$ where $\Eset$ is the set of undirected edges connecting nodes in $\Vset$. By the Hammersley-Clifford theorem~\citep{Clifford90}, any such distribution has the following form:
\begin{align}
	\Pr (\xvec \given \pgmpvec) =  \exp \bigg( \sum_{C \in \Cset} \pgmp_C \, T_C(\xvec_C) - A(\pgmpvec) \bigg)
	\label{EqnGraphicalModel}
\end{align}
where $\Cset$ is a set of cliques (fully-connected subgraphs) of $\graph$ and $T_C(\xvec_C)$ are the clique-wise sufficient statistics. For example, if $C = \{1,2,3\} \in \mathcal{C}$, then there would be a term $\pgmp_{1,2,3}\, T_{1,2,3}(\x_1,\x_2,\x_3)$ which involves the first, second and third random variables in $\xvec$.  Hence, a graphical model can be understood as an exponential family distribution with the form given in Eq.~\ref{EqnGraphicalModel}.  An important special case---which will be the focus in this paper---is a pairwise graphical model, where $\mathcal{C}$ consists of merely $\Vset$ and $\Eset$---i.e. $|C| = \{1,2\}, \forall C \in \mathcal{C}$, so that we have
\begin{align*}
	\Pr (\xvec \given \pgmpvec) =  \exp \bigg( \sum_{i \in \Vset} \pgmp_i T_{i}(x_i) + 
		\sum_{(i,j) \in \Eset} \pgmp_{ij} T_{ij}(x_i,x_j) - A(\pgmpvec)\bigg).
\end{align*}
Since graphical models provide direct interpretations on the Markov independence assumptions, for the Poisson-based graphical models in this section, we can easily investigate the conditional independence relationships between random variables rather than marginal correlations.

As an example, we will consider the Gaussian graphical model formulation of the standard multivariate normal distribution (for simplicity we will assume the mean vector is zero, i.e. $\bm{\mu} = 0$):
\begin{align}
\text{\bf Standard Form} & \quad \Leftrightarrow \quad \text{\bf Graphical Model Form} \notag\\
\Sigma = -\frac{1}{2}\Theta^{-1} & \quad \Leftrightarrow \quad\Theta = -\frac{1}{2}\Sigma^{-1} \label{eqn:ggm-parameter-transform}\\
\Pr(\xvec \given \Sigma) \propto \exp\bigg( -\frac{1}{2} \xvec^T\Sigma^{-1}\xvec \bigg) & \quad \Leftrightarrow \quad \Pr(\xvec \given \Theta) \propto \exp\bigg( \xvec^T \Theta \xvec \bigg) \notag \\
& \quad \Leftrightarrow \quad = \exp\bigg( \sum_{i} \theta_{ii} \x_i^2 + \sum_{i \neq j} \theta_{ij} \x_i\x_j \bigg) \, . \label{eqn:ggm-sum}
\end{align}
Note how Eq.~\ref{eqn:ggm-sum} is related to
Eq.~\ref{EqnGraphicalModel} by setting $\pgmp_i = \theta_{ii}$,
$\pgmp_{ij} = \theta_{ij}$, $T_i(\x_i) = x_i^2$, $T_{ij}(\x_i,\x_j) =
x_i x_j$ and $\Eset = \{ (i,j): i\neq j, \theta_{ij} \neq 0\}$---
i.e. the edges in the graph correspond to the non-zeros in $\Theta$.
In addition, this example shows that the marginal moments---i.e. the
covariance matrix $\Sigma$---are quite different from the graphical
model parameters---i.e. the negative of the inverse covariance matrix
$\Theta = -\frac{1}{2}\Sigma^{-1}$.  In general, for graphical models such as the Poisson graphical models defined in the next section, the transformation from the covariance to the graphical model parameter (Eq.~\ref{eqn:ggm-parameter-transform}) is not known in closed-form; in fact, this transformation is often very difficult to compute for non-Gaussian models \citep{wainwright2008graphical}.  For more information about graphical models and exponential families see \citep{koller2009probabilistic,wainwright2008graphical}.

\subsection{Poisson Graphical Model}

The first to consider multivariate extensions constructed by assuming conditional distributions are univariate exponential family distributions, such as and including the Poisson distribution, was \citet{Besag74}. In particular, suppose all node-conditional distributions---the conditional distribution of a node conditioned on the rest of the nodes---are univariate Poisson. Then, there is a unique joint distribution consistent with these node-conditional distributions \add{under some conditions}, and moreover this joint distribution is a graphical model distribution that factors according to a graph specified by the node-conditional distributions. In fact, this approach can be uniformly applicable for any exponential family beyond the Poisson distribution, and can be extended to more general graphical model settings~\citep{YRAL12,YRAL15} beyond the pairwise setting in \citep{Besag74}. The particular instance with the univariate Poisson as the exponential family underlying the node-conditional distributions is called a Poisson graphical model (PGM).\footnote{\citet{Besag74} originally named these Poisson auto models, focusing on pairwise graphical models, but here we consider the general graphical model setting.}

Specifically, suppose that for every $i \in \{1,\hdots, \nvar\}$, the node-conditional distribution is specified by univariate Poisson distribution in exponential family form as specified in Eq.~\ref{EqnUnivPoissExpFam}:
\begin{align}\label{EqnPGMNode}
	\Pr\big(\x_i \given \xvec_{-i} \big) = \exp \{ \condfunc(\xvec_{-i}) \, \x_i - \log (\x_i!) - \exp\big( \condfunc(\xvec_{-i}) \big) \} \, ,
\end{align}
where $\xvec_{-i}$ is the set of all $\x_j$ except $\x_i$, and the function $\condfunc(\xvec_{-i})$ is \emph{any} function that depends on the rest of all random variables except $\x_i$. Further suppose that the corresponding joint distribution on $\xvec$ factors according to the set of cliques $\Cset$ of a graph $\graph$.  \citet{YRAL15} then show that such a joint distribution consistent with the above node-conditional distributions exists, and moreover necessarily has the form
\begin{align}\label{EqnPGMGeneral}
	\Pr  (\xvec \given \bm{\pgmp}) = \exp\bigg\{ \sum_{\C \in \Cset} \pgmp_{\C} \prod_{i \in \C} \x_i - \sum_{i =1}^\nvar \log (\x_i!)  - A (\pgmpvec) \bigg\} \, ,
\end{align}
where the function $A(\pgmpvec)$ is the log-partition function on all parameters $\pgmpvec = \{\pgmp_{\C}\}_{\C \in \Cset}$. The pairwise PGM, as a special case, is defined as follows:
\begin{align}
	\PGM  (\xvec \given \pgmpvec) = \exp\bigg\{  \sum_{i =1}^\nvar \pgmp_i \x_i + \sum_{(i,j) \in \Eset} \pgmp_{ij} \x_i \x_j  - \sum_{i =1}^\nvar \log (\x_i!)  - A_{\text{PGM}}(\pgmpvec) \bigg\} \, ,
\end{align}
where $\Eset$ is the set of edges of the graphical model and $\pgmpvec = \{\pgmp_1, \pgmp_2, \cdots, \pgmp_\nvar \} \cup \{\pgmp_{ij}, \, \forall (i,j) \in \Eset \}$.  For notational simplicity and development of extensions to PGM, we will gather the node parameters $\pgmp_i$ into a vector $\nodepvec = [\pgmp_1, \pgmp_2, \cdots, \pgmp_\nvar] \in \R^\nvar$ and gather the edge parameters into a symmetric matrix $\edgepmat \in \R^{\nvar \times \nvar}$ such that $\edgep_{ij} = \edgep_{ji} = \pgmp_{ij}/2, \forall (i,j) \in \Eset$ and $\edgep_{ij} = 0, \forall (i,j) \not\in \Eset$.  Note that for PGM, $\edgepmat$ has \emph{zeros along the diagonal}. With this notation, the pairwise PGM can be equivalently represented in a compact vectorized form as:
\begin{align}
\PGM  (\xvec \given \nodepvec, \edgepmat ) &= \exp\{  \nodepvec^T \xvec + \xvec^T \edgepmat \xvec - {\textstyle \sum_{i=1}^\nvar} \log (\x_i!) - A_{\text{PGM}}(\nodepvec, \edgepmat) \} \label{EqnPGM} \, ,
\end{align}

Parameter estimation in a PGM is naturally suggested by its construction: all of the PGM parameters in Eq.~\ref{EqnPGM} can be estimated by considering the node-conditional distributions for each node separately, and solving an $\ell_1$-regularized Poisson regression for each variable. In contrast to the previous approaches in the sections above, this parameter estimation approach is not only simple, but is also guaranteed to be consistent even under high dimensional sampling regimes, under some other mild conditions including a sparse graph structural assumption (see \citet{YRAL12,YRAL15} for more details on the analysis).  As in Poisson log-normal models, the parameters of PGM can be made to depend on covariates to allow for more flexible correlations~ \citep{YRAL13b}.

In spite of its simple parameter estimation method, the major drawback with this vanilla Poisson graphical model distribution is that it only permits \emph{negative} conditional dependencies between variables:  
\begin{proposition}[\citet{Besag74}]\label{Prop:Negativity}
    Consider the Poisson graphical model distribution in Eq.~\ref{EqnPGM}. Then, for any parameters $\nodepvec$ and $\edgepmat$, $A_{\text{PGM}}(\nodepvec, \edgepmat) < +\infty$ only if the pairwise parameters are non-positive: $\edgep_{ij} \le 0, \, \forall (i,j) \in \Eset$ .
\end{proposition}
Intuitively, if any entry in $\edgepmat$, say $\edgepmat_{ij}$, is positive, the term $\edgepmat_{ij}\xvec_{i}\xvec_{j}$ in Eq.~\ref{EqnPGM} would grow quadratically, whereas the log base measure terms $-\log(\x_i!) - -\log(\x_j!)$ only decreases as $O(\x_i \log \x_i + \x_j \log \x_j)$, so $A(\nodepvec,\edgepmat) \to \infty$ as $\x_i,\x_j \to \infty$.  Thus, even though the Poisson graphical model is a natural extension of the univariate Poisson distribution (from the node-conditional viewpoint), it entails a highly restrictive parameter space, with severely limited applicability.  Thus, multiple PGM extensions attempt to relax this negativity restriction to permit positive dependencies as described next.

\subsection{Extensions of Poisson Graphical Models}
To circumvent the severe limitations of the PGM distribution 
which in particular only permits negative conditional dependencies, several extensions to PGM that permit a richer dependence structure have been proposed.

\subsubsection{Truncated PGM}
Because the negativity constraint is due in part to the infinite domain of count variable, a natural solution would be to truncate the domain of variables.
It was \citet{kaiser1997modeling} who first introduced an approach to truncate the Poisson distribution in the context of graphical models. Their idea was simply to use a Winsorized Poisson distribution for node-conditional distributions: $\x$ is a Winsorized Poisson if $z
= \mathbb{I}(z' < R) z' + \mathbb{I}(z' \geq R)R$, where $z'$ is Poisson, $\mathbb{I}(\cdot)$ is an indicator function, and $R$ is a fixed positive constant denoting the truncation level. However, \citet{YRAL13a} showed that Winsorized node-conditional distributions actually does \emph{not} lead to a consistent joint distribution. 

As an alternative way of truncation, \citet{YRAL13a} instead keep the same parametric form as PGM (Eq.~\ref{EqnPGM}) but merely truncate the domain to non-negative integers less than or equal to $R$---i.e. $\domain_{\text{TPGM}} = \{0,1,\cdots,R\}$, so that the joint distribution takes the form \citep{YRAL15}:
\begin{align}
\TPGM  (\xvec) &= \exp\{  \nodepvec^T \xvec + \xvec^T \edgepmat \xvec - {\textstyle \sum_i } \log (\x_i!) - {  A_{\text{TPGM}}}(\nodepvec, \edgepmat) \} \, .
\end{align}
As they show, the node-conditional distributions of this graphical model distribution belong to an exponential family that is Poisson-like, but with the domain bounded by $R$. Thus, the key difference from the vanilla Poisson graphical model (Eq.~\ref{EqnPGM}) is that the domain is finite, and hence the log partition function $A_{\text{TPGM}}(\cdot)$ only involves a finite number of summations. Thus, no restrictions are imposed on the parameters for the normalizability of the distribution.  

\citet{YRAL13a} discuss several major drawbacks to TPGM. First, the domain needs to be bounded a priori, so that $R$ should ideally be set larger than any unseen observation. Second, the effective range of parameter space for a non-degenerate distribution is still limited: as the truncation value $R$ increases, the effective values of pairwise parameters become increasingly negative or close to zero---otherwise, the distribution can be degenerate placing most of its probability mass at $0$ or $R$.

\subsubsection{Quadratic PGM and Sub-Linear PGM}
\citet{YRAL13a} also investigate the possibility of Poisson graphical models that (a) allows both positive and negative dependencies, as well as (b) allow the domain to range over all non-negative integers.  As described previously, a key reason for the negative constraint on the pairwise parameters $\edgep_{ij}$ in Eq.~\ref{EqnPGM} is that the log base measure $\sum_{i} \log (\x_i!)$ scales more slowly than the quadratic pairwise term $\xvec^T\edgepmat\xvec$ where $\xvec \in \Z_+^\nvar$. \citet{YRAL13a} thus propose two possible solutions: increase the base measure or decrease the quadratic pairwise term. 

First, if we modify the base measure of Poisson distribution with ``Gaussian-esque''  quadratic functions (note that for the linear sufficient statistics with positive dependencies, the base measures should be quadratic at the very least~\citep{YRAL13a}), then the joint distribution, which they call a quadratic PGM, is normalizable while allowing both positive and negative dependencies~\citep{YRAL13a}:
\begin{align}
\QPGM (\xvec) &= \exp\{  \nodepvec^T \xvec + \xvec^T {  \edgepmat } \xvec - A_{\text{QPGM}}(\nodepvec, {  \edgepmat } ) \}.
\end{align}
Essentially, QPGM has the same form as the Gaussian distribution, but where its domain is the set of non-negative integers. The key differences from PGM are that $\edgepmat$ can have negative values along the diagonal, and the Poisson base measure $\sum_i - \log(\x_i!)$ is replaced by the quadratic term $\sum_i \edgep_{ii} x_i^2$. Note that a sufficient condition for the distribution to be normalizable is given by: 
\begin{align} 
	\xvec^T \edgepmat \xvec  \,<\, -c\|\xvec\|_2^2 \quad \forall \xvec \in \Z_+^\nvar \, ,
\end{align}
for some constant $c > 0$, which in turn can be satisfied if $\edgepmat$ is negative definite. One significant drawback of QPGM is that the tail is Gaussian-esque and thin rather than  Poisson-esque and thicker as in PGM.

Another possible modification is to use sub linear sufficient statistics in order to preserve the Poisson base measure and possibly heavier tails. Consider the following univariate distribution over count-valued variables:
\begin{align}\label{EqnPoissonSL}
\Pr(z) \propto \exp\{ \theta T(z \,;\, R_0, R) - \log z! \} \, ,
\end{align}
which has the same base measure $\log z!$ as the Poisson, but with the following sub-linear sufficient statistics:
\begin{align}
    T(z \,;\, R_0, R) = \left\{ \begin{array}{l l}
                        z &\mbox{ if $z \leq R_0$} \\
                        -\frac{1}{2(R-R_0)}\ z^2 + \frac{R}{R-R_0}\ x -\frac{R_0^2}{2(R-R_0)} &\mbox{ if $R_0 < z \leq R$} \\
                        \frac{R+R_0}{2} &\mbox{ if $z \geq R$} \, .
                    \end{array} \right. \label{EqnSPGMSuff}
\end{align}
For values of $x$ up to $R_0$, $T(x)$ increases linearly, while after $R_0$ its slope decreases linearly, and finally after $R$, $T(x)$ becomes constant. The joint graphical model, which they call a sub-linear PGM (SPGM), specified by the node-conditional distributions belonging to the family in Eq.~\ref{EqnPoissonSL}, has the following form:
\begin{align}
&\SPGM  (\xvec) = \exp\{  \nodepvec^T {  T}(\xvec) + {  T}(\xvec)^T \, \edgepmat \, {  T}(\xvec) - {\textstyle \sum_i \log (\x_i!)} - A_{\text{SPGM}}(\nodepvec, \edgepmat \given {  R_0, R}) \} \, , 
\end{align}
where
\begin{align}
&A_{\text{SPGM}}(\nodepvec, \edgepmat \given {  R_0, R}) = \log \sum_{\xvec \in \Z_+} \exp\{  \nodepvec^T {  T}(\xvec) + {  T}(\xvec)^T \, \edgepmat \, {  T}(\xvec) - {\textstyle \sum_i \log (\x_i!)} \} \, ,
\end{align}
and $T(\xvec)$ is the entry-wise application of the function in Eq.~\ref{EqnSPGMSuff}. SPGM is always normalizable for $\edgep_{ij} \in \R \, \forall\, i\neq j$ \citep{YRAL13a}.

The main difficulty in estimating Poisson graphical model variants above with infinite domain is the lack of closed-form expressions for the log partition function, even just for the node-conditional distributions that are needed for parameter estimation. \citet{YRAL13a} propose an approximate estimation procedure that uses the univariate Poisson and Gaussian log partition functions as upper bounds for the node-conditional log-partition functions for the QPGM and SPGM models respectively.

\subsubsection{Poisson Square Root Graphical Model}
In the similar vein as SPGM in the earlier section, \citet{IRD16} consider the use of exponential families with square-root sufficient statistics. While they consider general graphical model families, their Poisson graphical model variant can be written as:
\begin{align}
\SQR(\xvec\given\theta) &= \exp\{ \nodepvec^T \sqrt{\xvec} + \sqrt{\xvec}^T \edgepmat \sqrt{\xvec} - {\textstyle \sum_i} \log (\x_i!) - A_{\text{SQR}}(\nodepvec, \edgepmat) \},
\end{align}
where $\edgep_{ii}$ can be non-zero in contrast to the zero diagonal of the parameter matrix in Eq.~\ref{EqnPGM}.  As with PGM, when there are no edges (i.e. $\edgep_{ij} = 0\,\,\forall i \neq j$) and $\nodepvec = 0$, this reduces to the independent Poisson model.  The node conditionals of this distribution have the form:
\begin{align}
\Pr(\x_i | \xvec_{-i} ) \propto \exp\big\{ \edgep_{ii} \x_i +  \big( \nodep_i + 2\edgepvec_{i,-i}^T \sqrt{\xvec_{-i}} \big) \sqrt{\x_i} - \log (\x_i!) \big\},
\end{align}
where $\edgepvec_{i,-i}$ is the $i$-th column of $\edgepmat$ with the the $i$-th entry removed.  This can be rewritten in the form of a \emph{two} parameter exponential family:
\begin{align}
\Pr(\x_i | \natp_1, \natp_2 ) = \exp\{ \natp_1 \x_i + \natp_2 \sqrt{\x_i} - \log (\x_i!) - A(\natp_1,\natp_2) \}\, , \label{EqnSQRNode}
\end{align}
where $\natp_1 = \edgep_{ii}$, $\,\natp_2 = \nodep_i + 2\edgepvec_{i,-i}^T \sqrt{\xvec_{-i}}$ and $A(\natp_1,\natp_2)$ is the log partition function.  Note that a key difference with the PGM variants in the previous section is that the diagonal of $\edgepmat_{\text{SQR}}$ can be non-zero whereas the diagonal of $\edgepmat_{\text{PGM}}$ must be zero.  Because the interaction term $\sqrt{\xvec}^T \edgepmat \sqrt{\xvec}$ is asymptotically linear rather than quadratic, the Poisson SQR graphical model does not suffer from the degenerate distributions of TPGM as well as the FLPGM discussed in the next section, while still allowing both positive and negative dependencies.

\begin{figure}[!ht]
\centering
\includegraphics[width=0.7\columnwidth]{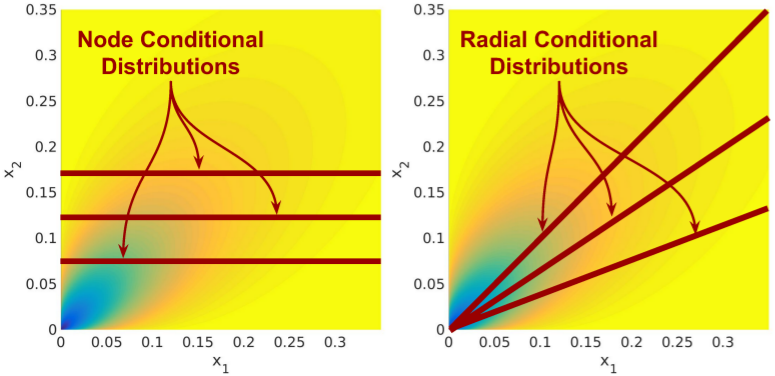}
\vspace{-1.8em}
\caption{\emph{Node}-conditional distributions (left) are univariate probability distributions of one variable conditioned on the other variables, while \emph{radial}-conditional distributions are univariate probability distributions of the vector scaling conditioned on the vector direction.  Both conditional distributions are helpful in understanding SQR graphical models. (Illustration from \citep{IRD16}.)}
\label{FigConditionals}
\end{figure}

To show that SQR graphical models can easily be normalized, \citet{IRD16} first define \emph{radial}-conditional distributions.  The \emph{radial}-conditional distribution assumes the \emph{unit direction} is fixed but the length of the vector is unknown.  The difference between the standard 1D \emph{node} conditional distributions and the 1D \emph{radial}-conditional distributions is illustrated in Fig.~\ref{FigConditionals}.  Suppose we condition on the unit direction $\uvec = \frac{\xvec}{\|\xvec\|_1}$ of the sufficient statistics but the scaling of this unit direction $\radx = \|\xvec\|_1$ is unknown. With this notation, \citet{IRD16} define the \emph{radial}-conditional distribution as:
\begin{align*}
\Pr(\xvec = \radx \uvec \given \uvec, \nodepvec, \edgepmat) &\propto \exp\{ \nodepvec^T\sqrt{\radx\uvec} + \sqrt{\radx\uvec}^T \edgepmat \sqrt{\radx\uvec} - {\textstyle \sum_i} \log((\radx\usca_i)!) \} \\
&\propto \exp\{ (\nodepvec^T \uvec) \sqrt{\radx} + (\sqrt{\uvec}^T \edgepmat \sqrt{\uvec}) \radx - {\textstyle \sum_i} \log((\radx\usca_i)!) \} \, .
\end{align*}
Similar to the node-conditional distribution, the radial-conditional distribution can be rewritten as a two parameter exponential family:
\begin{align}
&\Pr(\radx \given \uvec, \nodepvec, \edgepmat) = \exp\Big( \underbrace{\bar{\natp}_1\radx + \bar{\natp}_2 \sqrt{\radx}}_{O(\radx)}  + \underbrace{\tilde{B}_{\uvec}(\radx)}_{O(-\radx \log(\radx))} - A_{\text{rad}}(\bar{\natp}_1, \bar{\natp}_2) \Big) \label{eqn:radial-conditional} \, ,
 \end{align}
where $\bar{\natp}_1 = \sqrt{\uvec}^T \edgepmat \sqrt{\uvec}$, $\bar{\natp}_2 = \nodepvec^T \uvec$, and $\tilde{B}_{\uvec}(\radx) = -\sum_{i=1}^\nvar \log((\radx\usca_i)!)$.  The only difference between this exponential family and the node-conditional distribution is the different base measure---i.e. $-\sum_{i=1}^\nvar \log((\radx\usca_i)!) \neq -\log(\radx!)$. However, note that the log base measure is still $O(-\radx \log(\radx))$ and thus, the log base measure will overcome the linear term as $\radx \to \infty$.  Therefore, the \emph{radial}-conditional distribution is normalizable for \emph{any} $\bar{\natp}_1, \bar{\natp}_2 \in \R$.

With the radial-conditional distributions notation, \citet{IRD16} show that the log partition function for Poisson SQR graphical models is finite by separating the summation into a nested radial direction and scalar summation. Let $\Uset = \{ \uvec : \|\uvec\|_1 = 1, \uvec \in \R^\nvar \}$ be the set of unit vectors in the positive orthant. The SQR log partition function $A_{\text{SQR}}(\nodepvec,\edgepmat)$ can be decomposed into nested summation over the unit direction and the one dimensional radial conditional:
\begin{align}
A_{\text{SQR}}(\nodepvec, \edgepmat) &= \log \!\! \int\limits_{\uvec \in \Uset}  \sum\limits_{\radx \in \hat{\Z}} 
\exp\{ \bar{\natp}_1(\uvec \given \edgepmat) \,\radx + \bar{\natp}_2(\uvec \given \nodepvec) \sqrt{\radx} - {\textstyle \sum_i} \log( \radx \usca_i!) \} \mathrm{d}\uvec \, ,
\end{align}
where $\bar{\natp}_1(\uvec \given \edgepmat)$ and $\bar{\natp}_2(\uvec \given \nodepvec)$ are the radial conditional parameters as defined above and $\hat{\Z} = \{\radx: \radx\uvec \in \Z_+^\nvar \}$.  Note that $\hat{\Z} \subset \Z$, and thus the inner summation can be replaced by the radial-conditional log partition function.  Therefore, because $\Uset$ is a bounded set and the radial-conditional log partition function is finite for \emph{any} $\bar{\natp}_1(\uvec \given \nodepvec)$ and $\bar{\natp}_2(\uvec \given \edgepmat)$, $A_{\text{SQR}} < \infty$ and the Poisson SQR joint distribution is normalizable.

The main drawback to the Poisson SQR is that for parameter estimation, the log partition function $A(\natp_1,\natp_2)$ of the node conditionals in Eq.~\ref{EqnSQRNode} is not known in closed form in general.  \citet{IRD16} provide a closed-form estimate for the exponential SQR but a closed-form solution for the Poisson SQR model seems unlikely to exist.  \citet{IRD16} suggest numerically approximating $A(\natp_1,\natp_2)$, since it only requires a one dimensional summation.

\subsubsection{Local PGM}
Inspired by the neighborhood selection technique of \citet{meinshausen2006high}, \citet{allen2012log,allen2013local} propose to \remove{learn}\add{estimate} the network structure of count-valued data by fitting a series of $\ell_{1}$-regularized Poisson regressions to \remove{learn}\add{estimate} the node-neighborhoods.  Such an estimation method may yield interesting network estimates, but as \citet{allen2013local} note, these estimates do not correspond to a consistent joint density.  Instead, the underlying model is defined in terms of a series of local models where each variable is conditionally Poisson given its node-neighbors; this approach is thus termed the local Poisson graphical model (LPGM).  Note that LPGM does not impose any restrictions on the parameter space or types of dependencies; if the parameter space of each local model was constrained to be non-positive, then the LPGM reduces to the vanilla Poisson graphical model as previously discussed.  Hence, the LPGM is less interesting as a candidate multivariate model for count-valued data, but many may still find its simple and interpretable network estimates appealing.  Recently, several have proposed to adopt this estimation strategy for alternative network types \citep{hadiji2015poisson,han2016estimation}.

\subsubsection{Fixed-Length Poisson MRFs}
In a somewhat different direction, \citet{IRD15} propose a distribution that has the same parametric form as the original PGM, but allows positive dependencies by decomposing the joint distribution into two distributions.  The first distribution is the marginal distribution over the length of the vector denoted $\Pr(L)$---i.e. the distribution of the $\ell_1$-norm of the vector or the total sum of counts.  The second distribution, the fixed-length Poisson graphical model (FLPGM), is the conditional distribution of PGM \emph{given} the fact that the vector length $L$ is known or fixed, denoted $\FLPGM(\xvec \given \|\xvec\|_1 = L)$. Note that this allows the marginal distribution on length and the distribution given the length to be specified independently.\footnote{If the marginal distribution on the length is set to be the same as the marginal distribution on length for the PGM---i.e. if $\Pr(L) = \sum_{\xvec:\|\xvec\|_1 = L} \PGM(\xvec)$, then the PGM distribution is recovered.}  The restriction to negative dependencies is removed because the second distribution given the vector length $\FLPGM(\xvec \given \|\xvec\|_1 = L)$ has a finite domain {  $\domain_{\text{FLPGM}} = \{\xvec : \xvec \in \Z_+^d, \|\xvec\|_1 \!\!=\!\! L\} $} and is thus trivially normalizable---similar to the normalizability of the finite-domain TPGM.  More formally, \citet{IRD15} defined the FLPGM as:
\begin{align}
\Pr(\xvec \given \nodepvec, \edgepmat, \lambda) &= \Pr(L \given \lambda ) \,\, \FLPGM( \xvec \given {  \|\xvec\|_1 \!\!=\!\! L}, \nodepvec, \edgepmat) \, , \\
\FLPGM(\xvec \given {  \|\xvec\|_1 \!\!=\!\! L}, \nodepvec, \edgepmat) &= \exp\{  \nodepvec^T\xvec + \xvec^T \edgepmat \xvec - {\textstyle \sum_i } \log (\x_i!)  - {  A_L}(\nodepvec, \edgepmat) \} \, ,
\end{align}
where $\lambda$ is the parameter for the marginal length distribution---which could be Poisson, negative binomial or any other distribution on nonnegative integers.  In addition, FLPGM could be used as a replacement for the multinomial distribution because it has the same domain as the multinomial and actually reduces to the multinomial if there are no dependencies. \remove{In fact, FLPGM can be considered a multivariate generalization of the bivariate multiplicative binomial distribution defined in \citet{Altham1978}.}   \add{Earlier, \citet{Altham2012} developed an identical model by generalizing an earlier bivariate generalization of the binomial \citep{Altham1978}.  However, in \citep{Altham2012,Altham1978}, the model assumed that $L$ was constant over all samples, whereas \citep{IRD15} allowed for $L$ to vary for each sample according to some distribution $\Pr(L)$.}

One significant drawback is that FLPGM is not amenable to the simple node-wise parameter estimation method of the previous PGM models.  Nonetheless, in \citet{IRD15}, the parameters are heuristically estimated with Poisson regressions similar to PGM, though the theoretical properties of this heuristic estimate are unknown.  Another drawback is that while FLPGM allows for positive dependencies, the distribution can yet yield a degenerate distribution for large values of $L$---similar to the problem of TPGM where the mass is concentrated near 0 or $R$.  Thus, \citet{IRD15} introduce a decreasing weighting function $\omega(L)$ that scales the interaction term:
\begin{align}
&\FLPGM(\xvec \given \|\xvec\|_1 \!\!=\!\! L, \nodepvec, \edgepmat, {  \omega(\cdot)}) \notag \\
&\hspace{4em}= \exp\{  \nodepvec^T\xvec + { \omega(L)} \xvec^T \edgepmat \xvec - {\textstyle \sum_i } \log (\x_i!)  - A_L(\nodepvec, {  \omega(L)} \edgepmat) \} \, .
\end{align}

While the log-likelihood is not available in tractable form, \citet{IRD15} approximate the log likelihood using annealed importance sampling~\citep{Neal2001}, which might be applicable to the extensions covered previously as well.

\subsection{Summary of Conditional Poisson Generalizations}
The conditional Poisson models benefit from the rich literature in exponential families and undirected graphical models, or Markov Random Fields.  In addition, the conditional Poisson models have a simple parametric form.  The historical Poisson graphical model---or the auto-Poisson model \citep{Besag74})---only allowed negative dependencies between variables.  Multiple extensions have sought to overcome this severe limitation by altering the Poisson graphical model so that the log partition function is finite even with positive dependencies.  One major drawback to graphical model approach is that computing the likelihood requires approximation of the joint log partition function $A(\nodepvec, \edgepmat)$; a related problem is that the distribution moments and marginals are not known in closed-form.  Despite these drawbacks, parameter estimation using composite likelihood methods via $\ell_1$-penalized node-wise regressions (in which the joint likelihood is not computed) has solid theoretical properties under certain conditions.

\add{
\section{Model Comparison}
We compare models by first discussing two structural aspects of the models: (a) interpretability and (b) the relative stringency and ease of verifying theoretical assumptions and guarantees. We then present and discuss an empirical comparison of the models on three real-world datasets.

\subsection{Comparison of Model Interpretation}
{\bf Marginal models} can be interpreted as weakly decoupling modeling
marginal distributions over individual variables, from modeling the
dependency structure over the variables. However, in the discrete
case, specifically for distributions based on pairing copulas with
Poisson marginals, the dependency structure estimation is also
dependent on the marginal estimation, unlike for copulas paired with
continuous marginals \citep{genest2007primer}.  {\bf Conditional
models} or graphical models, on the other hand, can be interpreted as
specifying generative models for each variable given the variable's
neighborhood (i.e. the conditional distribution).  In addition,
dependencies in graphical models can be visualized and interpreted via
networks.  Here, each variable is a node and the weighted edges in the
network structure depict the pair-wise conditional dependencies
between variables.  The simple network depiction for graphical models may enable domain experts to interpret complex dependency structures more easily compared to other models.  Overall, marginal models may be preferred if modeling the statistics of the data, particularly the marginal statistics over individual variables, is of primary importance, while conditional models may be preferred if prediction of some variables given others is of primary importance.
{\bf Mixture models} may be more or less difficult to interpret
depending on whether there is an application-specific interpretation
of the latent mixing variable.  For example, a finite mixture of two
Poisson distributions may model the crime statistics of a city that
contains downtown and suburban areas.  On the other hand, a finite
mixture of fifty Poisson distributions or a log-normal Poisson mixture
when modeling crash severity counts (as seen in the empirical
comparison section) seems more difficult to interpret; even the model
empirically well fits the data, the hidden mixture variable might not have an obvious application-specific interpretation.

\subsection{Comparison of Theoretical Considerations}
Estimation of {\bf marginal models} from data has various theoretical
problems, as evidenced by the analysis of copulas paired with discrete
marginals in \citep{genest2007primer}.  The extent to which these
theoretical problems cause any significant practical issues remains
unclear. In particular, the estimators of the marginal distributions
themselves typically have easily checked assumptions since the empirical marginal distributions can be inspected directly.  On the other hand, the estimation of {\bf conditional models} is both computationally tractable, and comes with strong theoretical guarantees even under high-dimensional regimes where $\ninst < \nvar$ \citep{YRAL15}. However the assumptions under which the guarantees of the estimators hold are difficult to check in practice, and could cause problems if they are violated (e.g. outliers caused by unobserved factors). Estimation of {\bf mixture models} tend to have limited theoretical guarantees.  In particular, finite Poisson mixture models have very weak assumptions on the underlying distribution---eventually becoming a non-parametric distribution if $k = O(\ninst)$---but the estimation problems are likely NP-hard, with very few theoretical guarantees for practical estimators.  Yet, empirically as seen in the next section, estimating a finite mixture model using Expectation-Maximization iterations performs well in practice.
}

\subsection{Empirical Comparison}
\newcommand{\modellabel}[1]{``#1''}

In this section, we seek to empirically compare models from the three
classes presented to assess how well they fit real-world count data.

\add{
\subsubsection{Comparison Experimental Setup}
We empirically compare models on selected datasets from three diverse
domains which have different data characteristics in terms of their
mean count values and dispersion indices (Eq.~\ref{eqn:dispersion}) as
can be seen in Table~\ref{tab:datasets}. The crash severity dataset is
a small accident dataset from \citep{Milton2008} with three different
count variables corresponding to crash severity classes:
``Property-only'', ``Possible Injury'', and ``Injury''.  The crash
severity data exhibits high count values and high overdispersion.  We
retrieve raw next generation sequencing data for breast cancer (BRCA)
using the software TCGA2STAT \citep{Wan2016} and computed a simple
log-count transformation of the raw counts: $\lfloor \log(x+1)
\rfloor$, a common preprocessing technique for RNA-Seq data. The BRCA
data exhibits medium counts and medium overdispersion. We collect
the word count vectors from the Classic3 text corpus which contains
abstracts from aerospace engineering, medical and information sciences
journals.\footnote{\url{http://ir.dcs.gla.ac.uk/resources/test_collections/}}. The
Classic3 dataset exhibits low counts---including many zeros---and
medium overdispersion. 
In the supplementary material, we also give results for a crime
statistics dataset and the 20 Newsgroup dataset but they have similar
characteristics and perform similarly to the the BRCA and Classic3
datasets respectively; thus, we omit them for simplicity. We
select variables (e.g. for $\nvar = 10$ or $\nvar=100$) by sorting
the variables by mean count value---or sorting by variance in the case
of the BRCA dataset as highly variable genes are of more interest in biology.

\begin{table}
\centering
\caption{Dataset Statistics}\label{tab:datasets}
\vspace{0.5em}
\resizebox{1\textwidth}{!}{
\begin{tabular}{llllllllllllll}
\toprule
\multicolumn{3}{r}{(Per Variable $\Rightarrow$)} & \multicolumn{3}{c}{Means} & \multicolumn{3}{c}{Dispersion Indices} & 
  \multicolumn{3}{c}{Spearman's $\rho$} \\
\cmidrule(r){4-6} \cmidrule(r){7-9} \cmidrule(r){10-12}
Dataset & $\nvar$ & $\ninst$ & Min & Med & Max & Min & Med & Max & Min & Med & Max \\
\midrule
Crash Severity & 3 & 275 & 3.4 & 3.8 & 9.7  & 6 & 9.3 & 16  & 0.61 & 0.73 & 0.79  \\
BRCA & 10 & 878 & 3.2 & 5 & 7.7  & 1.5 & 2.2 & 3.8  & -0.2 & 0.25 & 0.95  \\
 & 100 & 878 & 1.1 & 4 & 9  & 0.63 & 1.7 & 4.6  & -0.5 & 0.08 & 0.95  \\
 & 1000 & 878 & 0.51 & 3.5 & 11  & 0.26 & 1 & 4.6  & -0.64 & 0.06 & 0.97  \\
Classic3 & 10 & 3893 & 0.26 & 0.33 & 0.51  & 1.4 & 3.4 & 3.8  & -0.17 & 0.12 & 0.82  \\
 & 100 & 3893 & 0.09 & 0.14 & 0.51  & 1.1 & 2.1 & 8.3  & -0.17 & 0.02 & 0.82  \\
 & 1000 & 3893 & 0.02 & 0.03 & 0.51  & 0.98 & 1.7 & 8.5  & -0.17 & -0 & 0.82  \\
\bottomrule
\end{tabular}
}
\end{table}

In order to understand how each model might perform under varying data
characteristics, we consider the following two questions: (1) How well
does the model (i.e. the joint distribution) fit the underlying data
distribution? (2) How well does the model capture the dependency
structure between variables?  To help answer these questions, we
evaluate the empirical fit of models using two metrics, which only 
require samples from the model.  The
first metric is based on a statistic called maximum mean discrepancy
(MMD) \citep{Gretton2012} which estimates the maximum moment
difference over all possible moments.  The empirical MMD can be
approximated as follows from two sets of samples $X \in \R^{\ninst_1
  \times \nvar}$ and $Y \in \R^{\ninst_2 \times \nvar}$: 
\begin{align}
\widehat{\text{MMD}}(\mathcal{G}, X, Y) = \sup_{f\in\mathcal{G}} \,\, \frac{1}{\ninst_1} \sum_{i=1}^{\ninst_1} f(\bm{x}_i) - \frac{1}{\ninst_2} \sum_{j=1}^{\ninst_2} f(\bm{y}_j) \, , 
\end{align}
where $\mathcal{G}$ is the union of the RKHS spaces based on the Gaussian kernel using twenty one $\sigma$ values log-spaced between 0.01 and 100. In our experiments, we estimate the MMD between the pairwise marginals of model samples and the pairwise marginals of the original observations:
\begin{align}
D^{\text{MMD}}_{st} = \left\{ 
\begin{array}{ll} \widehat{\text{MMD}}(\mathcal{G}, [\xvec^{(s)}],\, [\widehat{\xvec}^{(s)}] ), & s = t \\
 \widehat{\text{MMD}}(\mathcal{G}, [\xvec^{(s)}, \xvec^{(t)}],\, [\widehat{\xvec}^{(s)}, \widehat{\xvec}^{(t)}] ), & \text{otherwise} 
\end{array} \right. \, .
\end{align}
where \add{$\xvec^{(s)}$} is the vector of data for the \add{$s$-th} variable of the true data and \add{$\widehat{\xvec}^{(s)}$} is the vector of data for the \add{$s$-th} variable of samples from the estimated model---i.e. \add{$\xvec^{(s)}$} are observations from the true underlying distribution and \add{$\widehat{\xvec}^{(s)}$} are samples from the estimated model distribution. In our experiments, we use the fast approximation code for MMD from \citep{Zhao2015} with $2^6$ number of basis vectors for the FastMMD approximation algorithm. The second metric merely computes the absolute difference between the pairwise Spearman's $\rho$ values of model samples and the Spearman's $\rho$ values of the original observations:
\begin{align}
D^{\rho}_{st} = |\rho(\xvec^{(s)}, \xvec^{(t)}) - \rho(\widehat{\xvec}^{(s)}, \widehat{\xvec}^{(t)})|, \quad \forall s, t \, .
\end{align}
The MMD metric is of more general interest because it evaluates
whether the models actually fit the empirical data distribution while the Spearman metric may be more interesting for practitioners who primarily care about the dependency structure, such as biologists who specifically want to study gene dependencies rather than gene distributions.

We empirically compare the model fits on these real-world data sets
for several types of models from the three general classes presented.
As a baseline, we estimate an independent 
Poisson model (\modellabel{Ind Poisson}). We include Gaussian copulas
and vine copulas both paired with Poisson marginals
(\modellabel{Copula Poisson} and \modellabel{Vine Poisson}) to
represent the marginal model class.  We estimate the copula-based
models via the two-stage Inference Functions for Margins (IFM) method
\citep{Joe1996} via the distributional transform
\citep{Ruschendorf2013}. For the mixture class, we include both a
simple finite mixture of independent Poissons (\modellabel{Mixture
  Poiss}) and a log-normal mixture of Poissons
(\modellabel{Log-Normal}).  The finite mixture was estimated using a
simple expectation-maximization (EM) algorithm; the log-normal mixture
model was estimated via MCMC sampling using the code from
\citep{Zhan2015}. For the conditional model class, we estimate the
simple Poisson graphical model (\modellabel{PGM}), which only allows
negative dependencies, and three variants that allow for positive
dependencies: the truncated Poisson graphical model
(\modellabel{Truncated PGM}), the Fixed-Length Poisson graphical model
with a Poisson distribution on the vector length $L = \|x\|_1$
(\modellabel{FLPGM Poisson}) and the Poisson square root graphical
model (\modellabel{Poisson SQR}). Using composite likelihood methods of penalized $\ell_1$ node-wise regressions, we estimate these models via code from \citep{YRAL15}, \citep{IRD14a},
\citep{Inouye2016a} and the
\texttt{XMRF}\footnote{\url{https://cran.r-project.org/web/packages/XMRF/index.html}}
R package.  After parameter estimation, we generate 1,000 samples for
each method using different types of sampling for each of the model
classes. 

To avoid overfitting to the data, we employ 3-fold cross-validation
and report the average over the three 
folds.  Because the conditional models (PGM, TPGM, FLPGM, and Poisson
SQR) can be significantly different depending on the regularization
parameter---i.e. the weight for the $\ell_1$ regularization term in
the objective function for these models---, we select the
regularization parameter of these models by computing the metrics on a
tuning split of the training data. For the mixture model, we similarly
tune the number of components $k$ by testing $k = \{10,20,30,\cdots,
100\}$.  For the very high dimensional datasets where $\nvar = 1000$,
we use a regularization parameter near the tuning parameters found
when $\nvar = 100$ and fix $k = 50$ in order to avoid the extra
computation of selecting a parameter. More sampling and implementation
details for each model are available in the supplementary material. 
}

\add{
\subsubsection{Empirical Comparison Results}
The full results for both the MMD and Spearman's $\rho$ metrics for the
crash severity, breast cancer RNA-Seq and Classic3 text datasets can
be seen in Fig.~\ref{fig:crash-severity}, Fig.~\ref{fig:brca}, and
Fig.~\ref{fig:classic3} respectively.  The low dimensional results
($\nvar \leq 10$) give evidence across all the datasets that three
models outperform the others in their classes:\footnote{\add{For the
    crash-severity dataset, the truncated Poisson graphical model
    (\modellabel{Truncated PGM}) outperforms the Poisson SQR model
    under the pairwise MMD metric.  After inspection, however, we
    realized that the Truncated PGM model performed better merely
    because outlier values were truncated to the 99th percentile as
    described in the supplementary material. This reduced the overfitting of outlier
    values caused by the crash severity dataset's high
    overdispersion.}}  
The Gaussian copula paired with Poisson marginals model
(\modellabel{Copula Poisson}) for the marginal model class, the
mixture of Poissons distribution (\modellabel{Mixture Poiss}) for the
mixture model class, and the Poisson SQR distribution
(\modellabel{Poisson SQR}) for the conditional model class. 
Thus, we only include these representative models along with an independent Poisson baseline in the high-dimensional experiments when $\nvar > 10$. We discuss the results for specific data characteristics as represented by each dataset.\footnote{\add{These basic trends are also corroborated by the two datasets in the supplementary material.}}

For the crash severity dataset with high counts and high
overdispersion (Fig.~\ref{fig:crash-severity}), mixture models
(i.e. \modellabel{Log-Normal} and \modellabel{Mixture Poiss}) perform
the best as expected since they can model overdispersion well.
However, if dependency structure is the only object of interest, the
Gaussian copula paired with Poisson marginals
(\modellabel{Copula Poisson}) performs well. For the BRCA dataset with
medium counts and medium overdispersion (Fig.~\ref{fig:brca}), we note
similar trends with two notable exceptions: (1) The Poisson SQR model
actually performs reasonably in low dimensions suggesting that it can
model moderate overdispersion.  (2) The high dimensional ($d \geq
100$) Spearman's $\rho$ difference results show that the Gaussian
copula paired with Poisson marginals (\modellabel{Copula
  Poisson}) performs significantly better than the mixture model; this
result suggests that copulas paired with Poisson marginals are likely
better for modeling dependencies than mixture models. Finally, for the
Classic3 dataset with low counts and medium overdispersion
(Fig.~\ref{fig:classic3}), the Poisson SQR model seems to perform well
in this low-counts setting especially in low dimensions unlike in
previous data settings.  While the simple independent mixture of
Poisson distributions still performs well, the Poisson log-normal
mixture distribution (\modellabel{Log-Normal}) performs quite poorly
in this setting with small counts and many zeros. This poor
performance of the Poisson log-normal mixture is somewhat surprising
since the dispersion indices are almost all greater than one as seen
in Table~\ref{tab:datasets}.  The differing results between low counts
and medium counts with similar overdispersion demonstrate the
importance to consider both the overdispersion and the mean count
values when characterizing a dataset. 

In summary, we note several overall trends. Mixture models are
important for overdispersion when counts are medium or high. The
Gaussian copula with Poisson marginals joint distribution can estimate
dependency structure (per the Spearman metric) for a wide range of
data characteristics even when the distribution does not fit the
underlying data (per the MMD metric). The Poisson SQR model performs
well for low count values with many zeros (i.e. sparse data) and may
be able to handle moderate overdispersion. 

\begin{figure}[!ht]
\newcommand{\histwidth}{0.32\textwidth}
\centering
\includegraphics[width=0.6\textwidth]{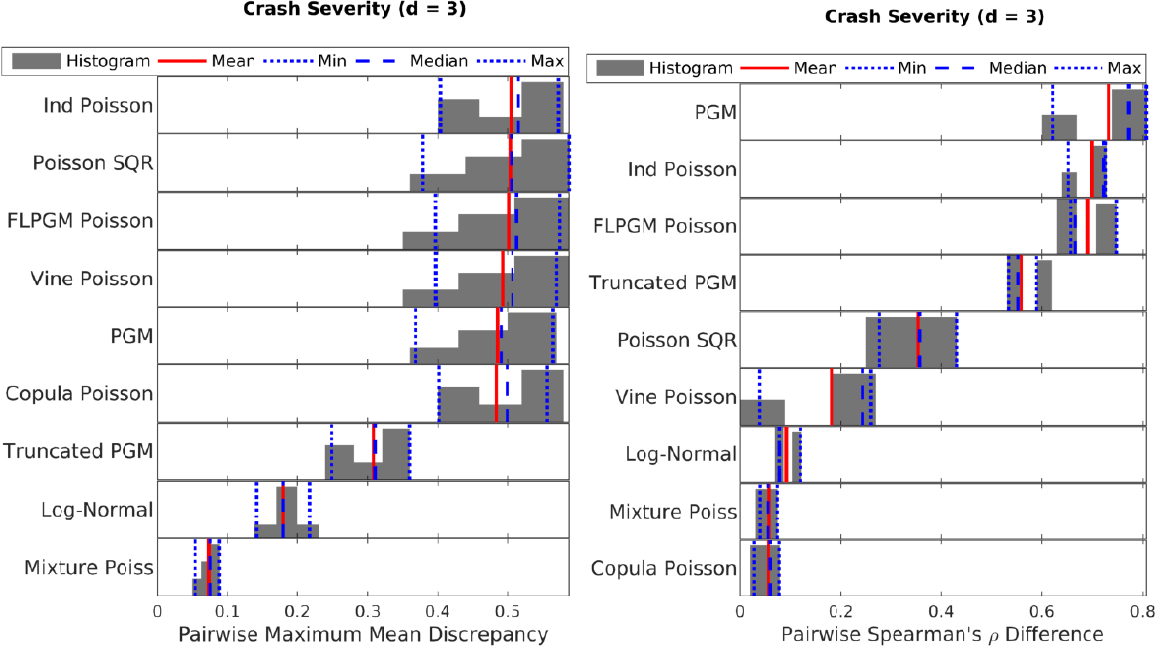}
\caption{\add{Crash severity dataset (high counts and high overdispersion): MMD (left) and Spearman $\rho$'s difference (right). As expected, for high overdispersion, mixture models (\modellabel{Log-Normal} and \modellabel{Mixture Poiss}) seem to perform the best.}}
\label{fig:crash-severity}
\end{figure}

\begin{figure}[!ht]
\newcommand{\histwidth}{0.32\textwidth}
\centering
\includegraphics[width=\textwidth]{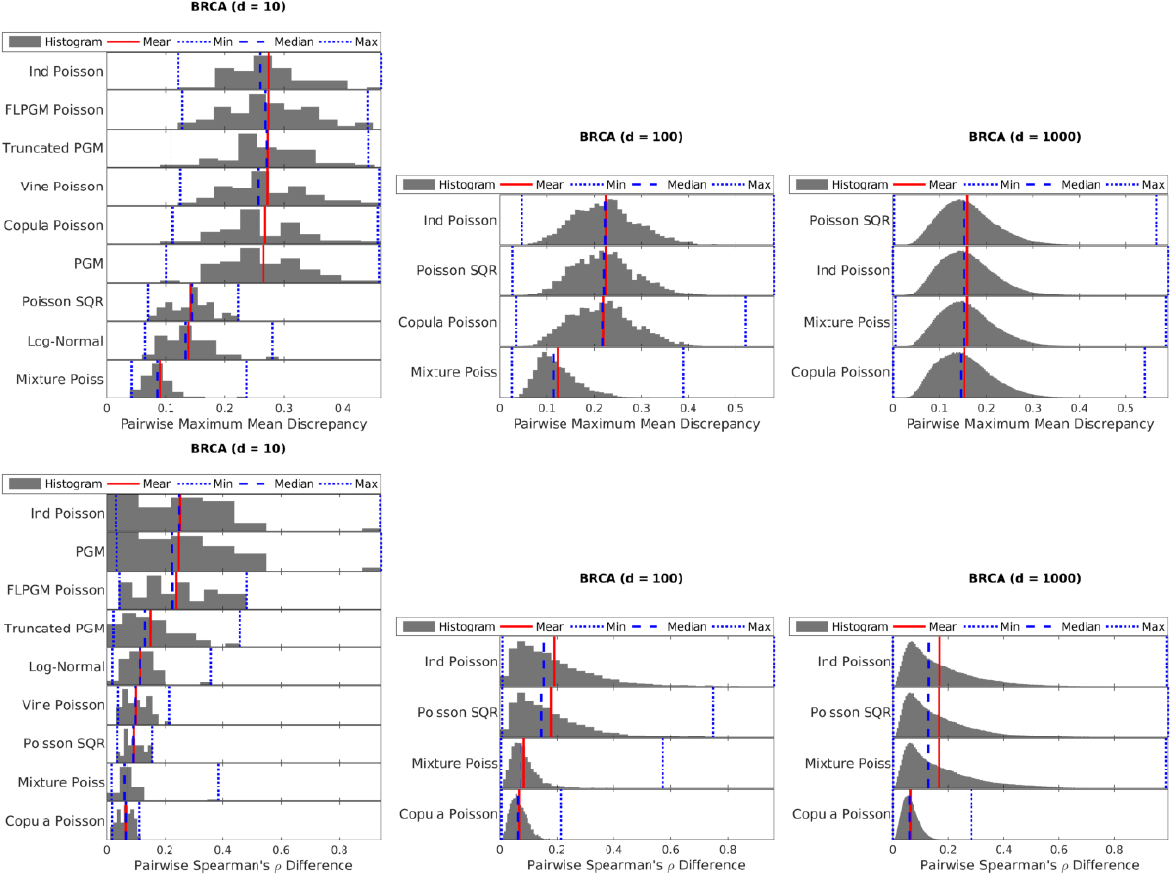}
\caption{\add{BRCA RNA-Seq dataset (medium counts and medium overdispersion): MMD (top) and Spearman $\rho$'s difference (bottom) with different number of variables: 10 (left), 100 (middle), 1000 (right). While mixtures (\modellabel{Log-Normal} and \modellabel{Mixture Poiss}) perform well in terms of MMD, the Gaussian copula paired with Poisson marginals (\modellabel{Copula Poisson}) can model dependency structure well as evidenced by the Spearman metric.}}
\label{fig:brca}
\end{figure}

\begin{figure}[!ht]
\newcommand{\histwidth}{0.32\textwidth}
\centering
\includegraphics[width=\textwidth]{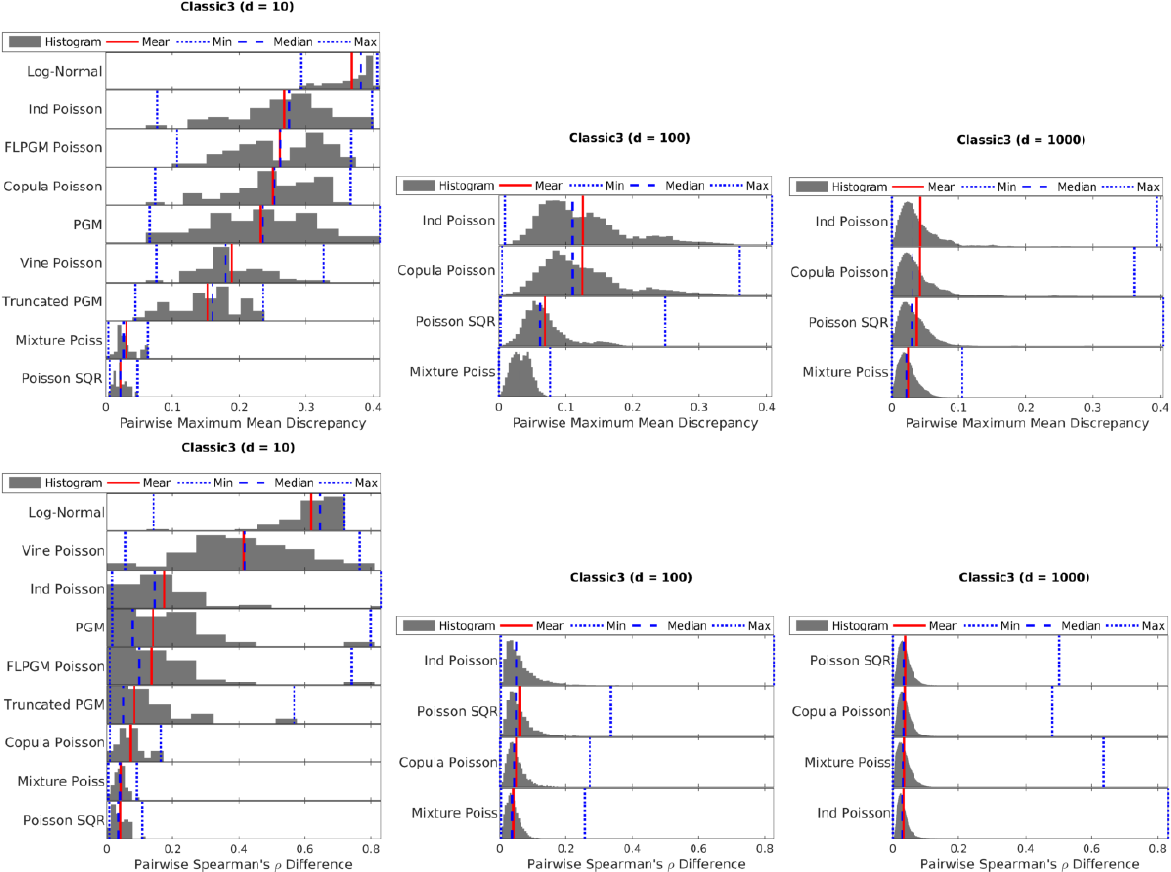}
\caption{\add{Classic3 text dataset (low counts and medium overdispersion): MMD (top) and Spearman $\rho$'s difference (bottom) with different number of variables: 10 (left), 100 (middle), 1000 (right).  The Poisson SQR model performs better on this low count dataset than in previous settings.}}
\label{fig:classic3}
\end{figure}
}

\clearpage

\newcommand{\captionup}{\vspace{-1em}}

\section{Discussion}
While this review analyzes each model class separately, it would be quite interesting to consider combinations or synergies between the model classes.  \add{Because negative binomial distributions can be viewed as a gamma-Poisson mixture model, one simple idea is to consider pairing a copula with negative binomial marginals or developing a negative binomial SQR graphical model.}  \remove{One such idea that we have implicitly explored is the combination of a copula model with the negative binomial---which can be viewed as a gamma-Poisson mixture model.}  As another example, we could form a finite mixture of copula-based or graphical-model-based models.  This might combine the strengths of a mixture in handling multiple modes and overdispersion with the strengths of the copula\add{-based models} and graphical models which can explicitly model dependencies.  \remove{As another example, because the negative binomial models perform much better for the non-text datasets, we might consider how to construct a negative binomial graphical model such as a negative binomial SQR model instead of a Poisson SQR model.}

We may also consider how one type of model informs the other.  For example, by the generalized Sklar's theorem \citep{Ruschendorf2013}, each conditional Poisson \add{graphical} model actually induces a copula---just as the Gaussian graphical model induces the Gaussian copula.  Studying the copulas induced by graphical models seems to be a relatively unexplored area.  On the other side, it may be useful to consider fitting a Gaussian copula \add{paired with discrete marginals} using the theoretically-grounded techniques from graphical models for sparse dependency structure estimation especially for the small sample regimes in which $\nvar > \ninst$; this has been studied for the case of continuous marginals in \citep{Liu2012a}.  Overall, bringing together and comparing these diverse paradigms for probability models opens up the door for many combinations and synergies.
 
\section{Conclusion}

We have reviewed three main approaches to constructing
multivariate distributions derived from the Poisson using three different assumptions: 1) the marginal distributions are derived from the Poisson, 2) the joint distribution is a mixture of independent Poisson distributions, and 3) the node-conditional distributions are derived from the Poisson. The first class based on Poisson marginals, and in particular the general \add{approach of pairing copulas with Poisson marginals,} \remove{copula approach,} provides an elegant way to \add{partially}\footnote{\add{In the discrete case, the dependency structure cannot be perfectly decoupled from the marginal distributions unlike in the continuous case where the dependency structure and marginals can be perfectly decoupled.}} decouple the marginals from the dependency structure and gives \remove{excellent} \add{strong} empirical results despite some theoretical issues related to non-uniqueness.  While advanced \remove{copula} methods \add{to estimate the joint distribution of copulas paired with discrete marginals} such as simulated likelihood \citep{Nikoloulopoulos2016} or vine copula constructions provide more accurate or more flexible copula models respectively, our empirical results suggest that a simple Gaussian copula \add{paired with Poisson marginals} \remove{model} with the trivial distributional transform (DT) can perform quite well in practice.  The second class based on mixture models can be particularly helpful for handling overdispersion that often occurs in real count data with the log-normal-Poisson mixture \add{and a finite mixture of independent Poisson distributions} \remove{ and negative binomial---a.k.a. gamma-Poisson mixture---} being prime examples.  In addition, mixture models have closed-form moments and in the case of a finite mixture, closed-form likelihood calculations---something not generally true for the other classes.  The third class based on Poisson conditionals can be represented as graphical models, thus providing both compact and visually appealing representations of joint distributions.  Conditional models benefit from strong theoretical guarantees about model recovery given certain modeling assumptions. However, \remove{ our experiments suggest that the modeling assumptions required for the conditional models} \add{checking conditional modeling assumptions may be impossible and} may not always be satisfied for real-world count data.
\add{From our empirical experiments, we found that (1) mixture models are important for overdispersion when counts are medium or high, (2) the Gaussian copula with Poisson marginals joint distribution can estimate dependency structure for a wide range of data characteristics even when the distribution does not fit the underlying data, and (3) Poisson SQR models perform well for low count values with many zeros (i.e. sparse data) and can handle moderate overdispersion.} Overall, in practice, we would recommend comparing the three best performing methods from each class: namely the Gaussian copula model \add{paired with Poisson marginals,} \remove{with negative binomial marginals,} the \add{finite mixture of independent Poisson distributions}, \remove{ Poisson mixture model} and the Poisson SQR model. This initial comparison will likely highlight some interesting properties of a given dataset and suggest which class to pursue in more detail.

This review has highlighted several key strengths and weaknesses of
the main approaches to constructing multivariate Poisson
distributions.  Yet, there remain many open questions.  For example, what are the marginal distributions of the Poisson graphical models which are defined in terms of their conditional distributions?  Or conversely, what are the conditional distributions of the copula models which are defined in terms of their marginal distributions?  Can novel models be created at the intersection of these model classes that could combine the strengths of different classes as suggested in the discussion section?  Could certain model classes be developed in an application area that has been largely dominated by another model class?  For example, graphical models are well-known in the machine learning literature while copula models are well-known in the financial modeling literature.
Overall, multivariate Poisson models are poised to increase in popularity given
the wide potential applications to real-world high-dimensional
count-valued data in text analysis, genomics, spatial statistics,
economics, and epidemiology.

\subsection*{Acknowledgments}
D.I. and P.R. acknowledge the support of ARO via W911NF-12-1-0390 and NSF
via IIS-1149803, IIS-1447574, DMS-1264033, and NIH via R01 GM117594-01
as part of the Joint DMS/NIGMS Initiative to Support Research at the
Interface of the Biological and Mathematical
Sciences. G.A. acknowledges support from NSF DMS-1264058 and NSF DMS-1554821.

\bibliographystyle{abbrvnat}
\bibliography{mult-poisson-review}

\clearpage
\appendix
\section{Supplementary Datasets and Results}
\begin{figure}[!ht]
\newcommand{\histwidth}{0.32\textwidth}
\centering
\includegraphics[width=\histwidth]{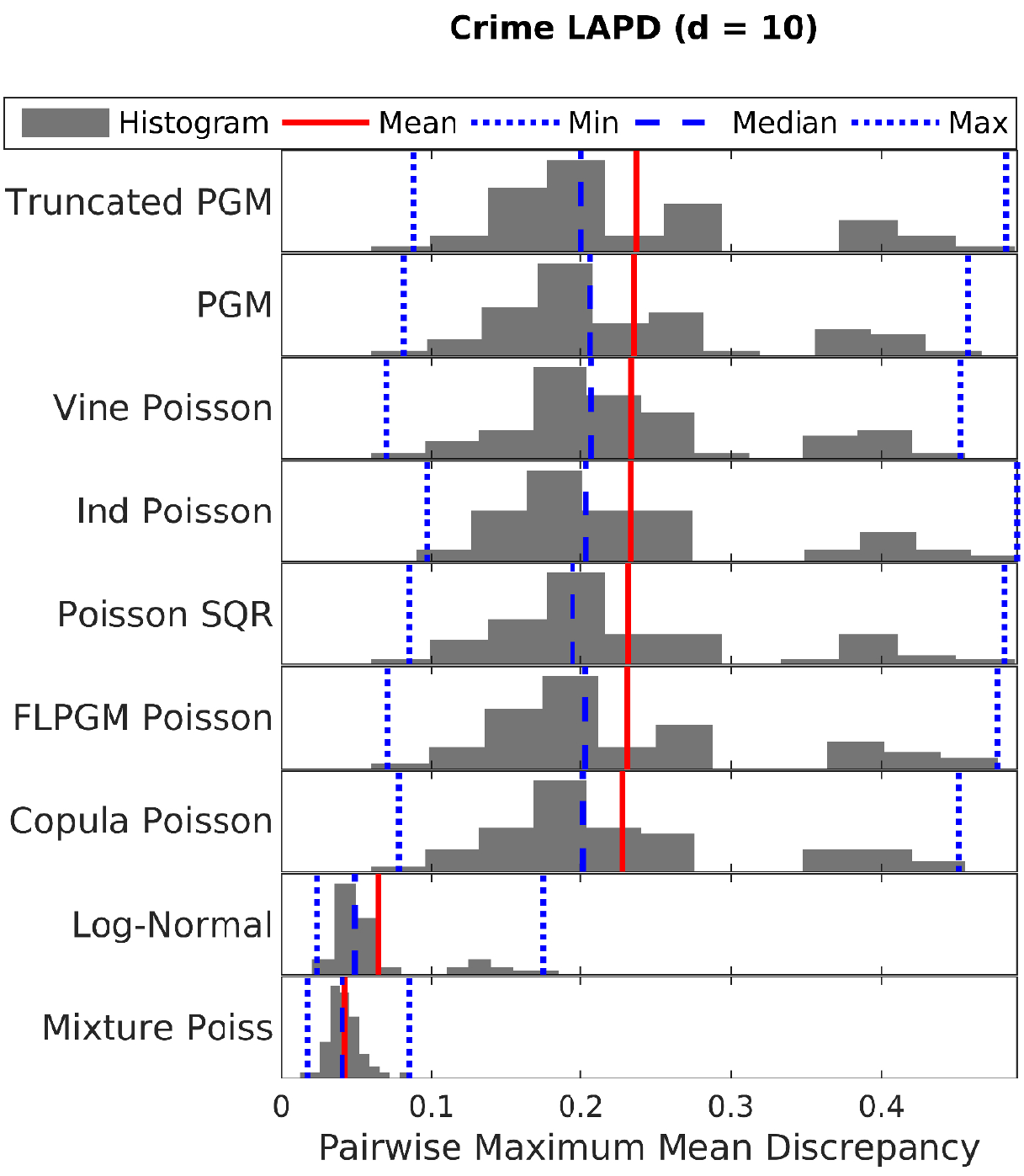}
\includegraphics[width=\histwidth]{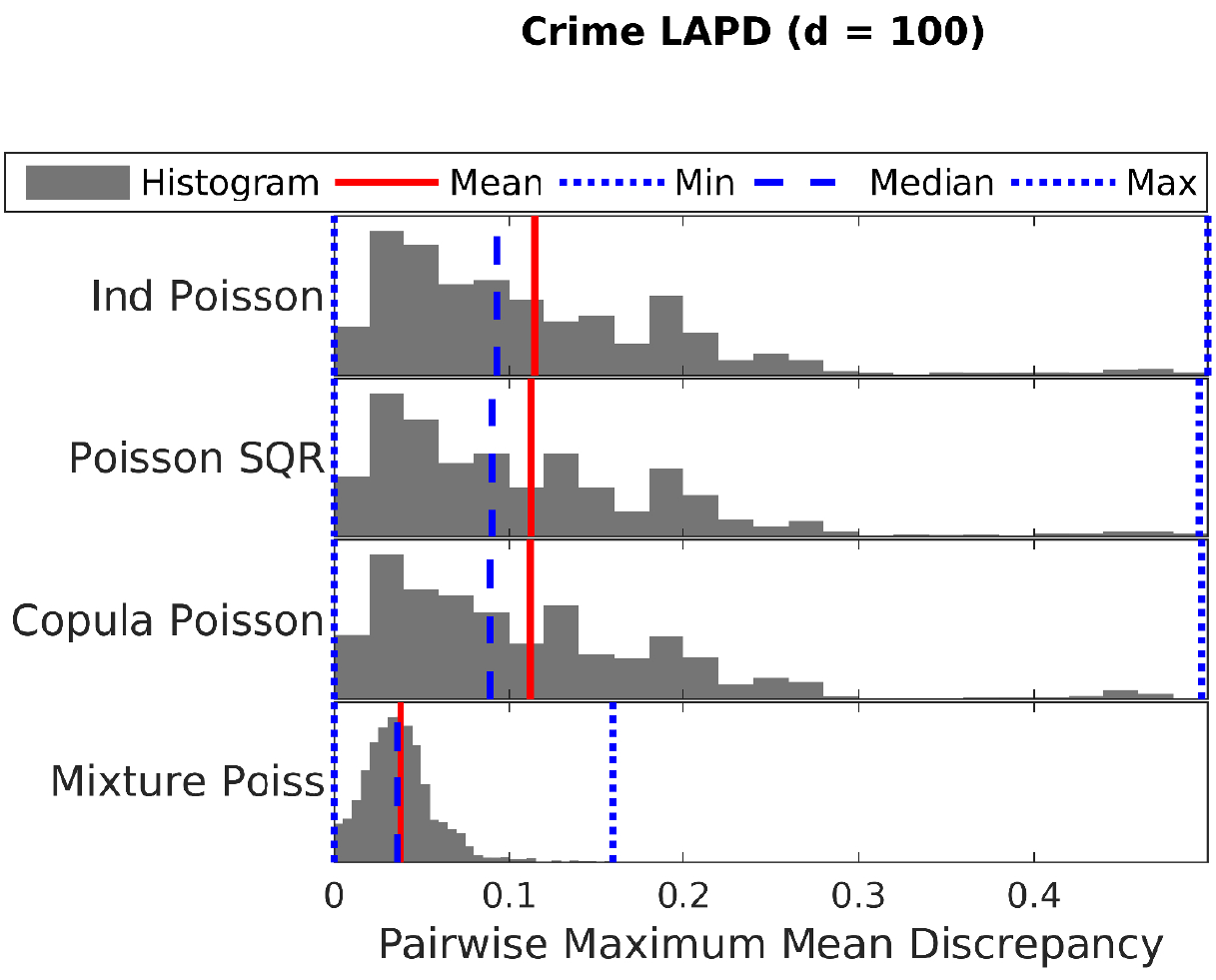} \\
\includegraphics[width=\histwidth]{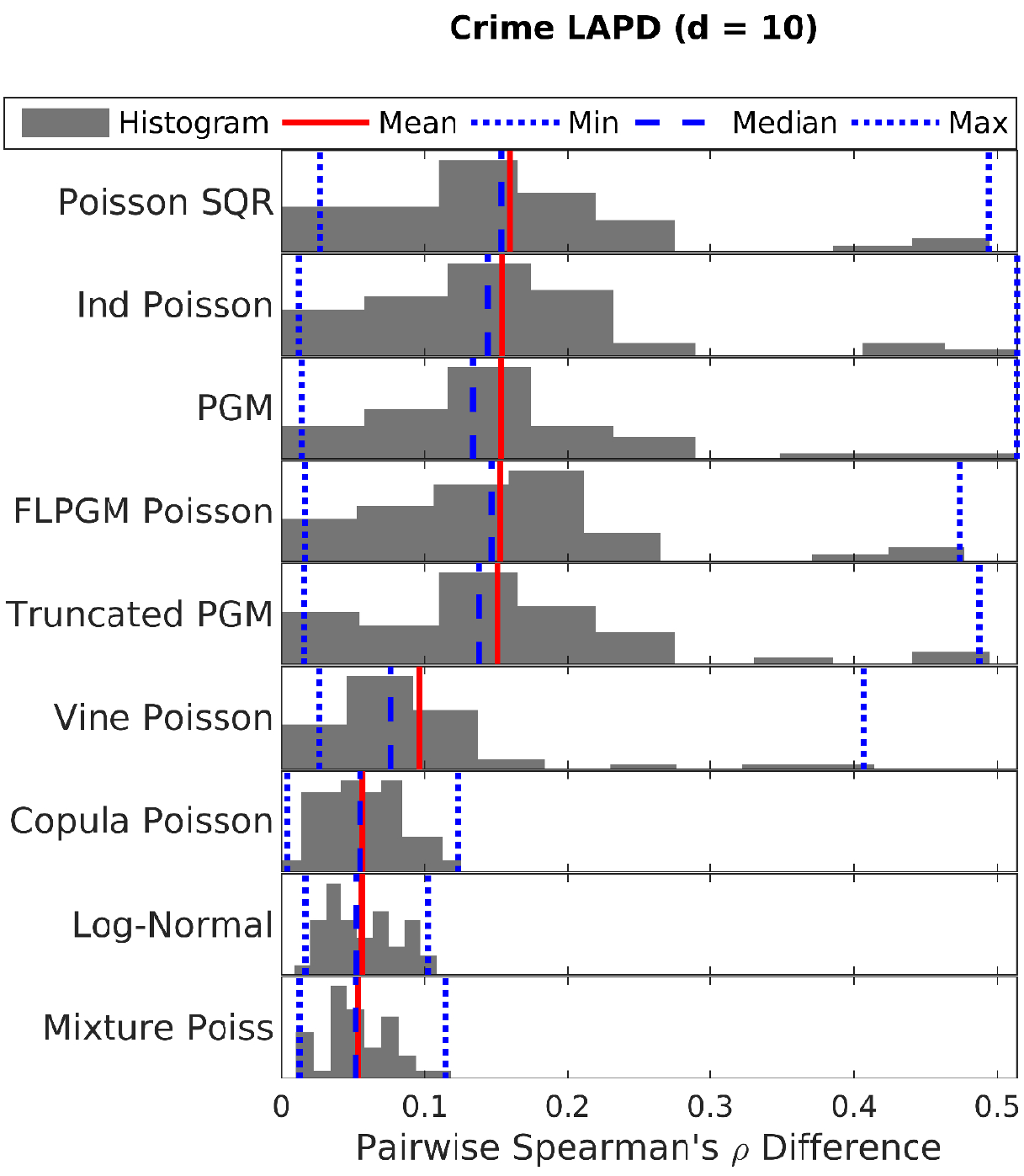}
\includegraphics[width=\histwidth]{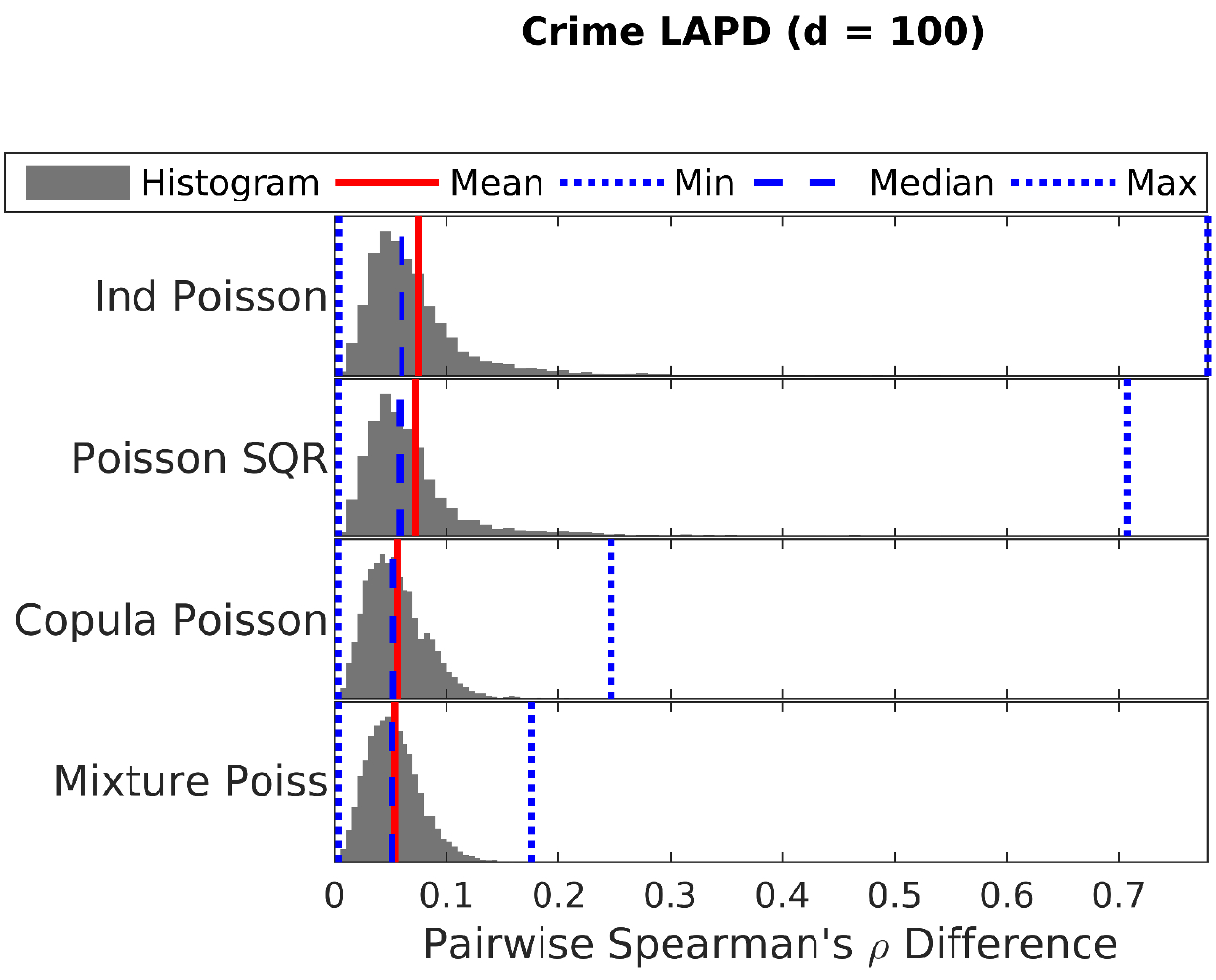}
\caption{The results for the LAPD crime statistics dataset with medium count values and medium overdispersion behave similarly to the results from the BRCA dataset described in the paper.}
\label{fig:crime-lapd}
\end{figure}

\begin{figure}[!ht]
\newcommand{\histwidth}{0.32\textwidth}
\centering
\includegraphics[width=\histwidth]{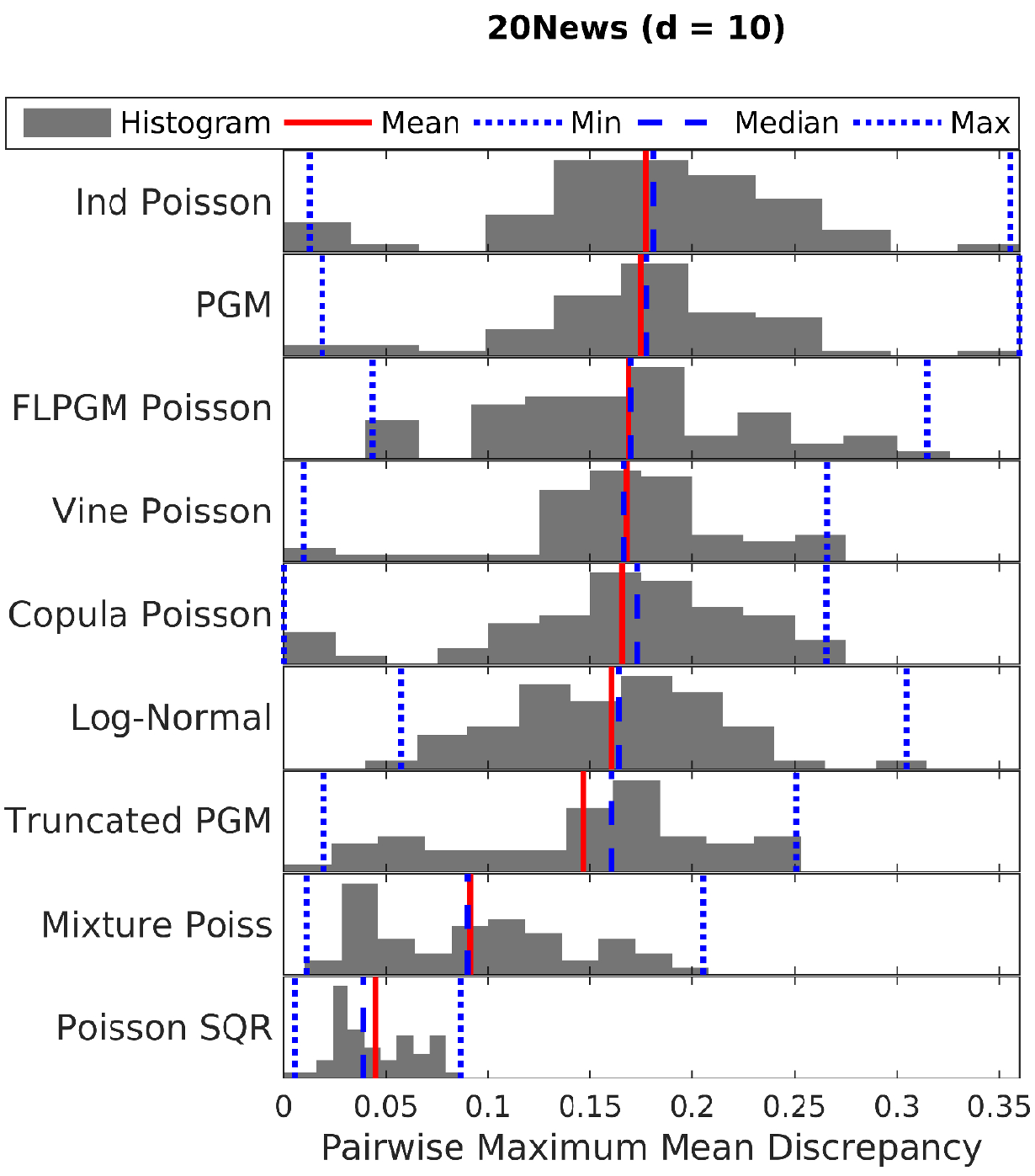}
\includegraphics[width=\histwidth]{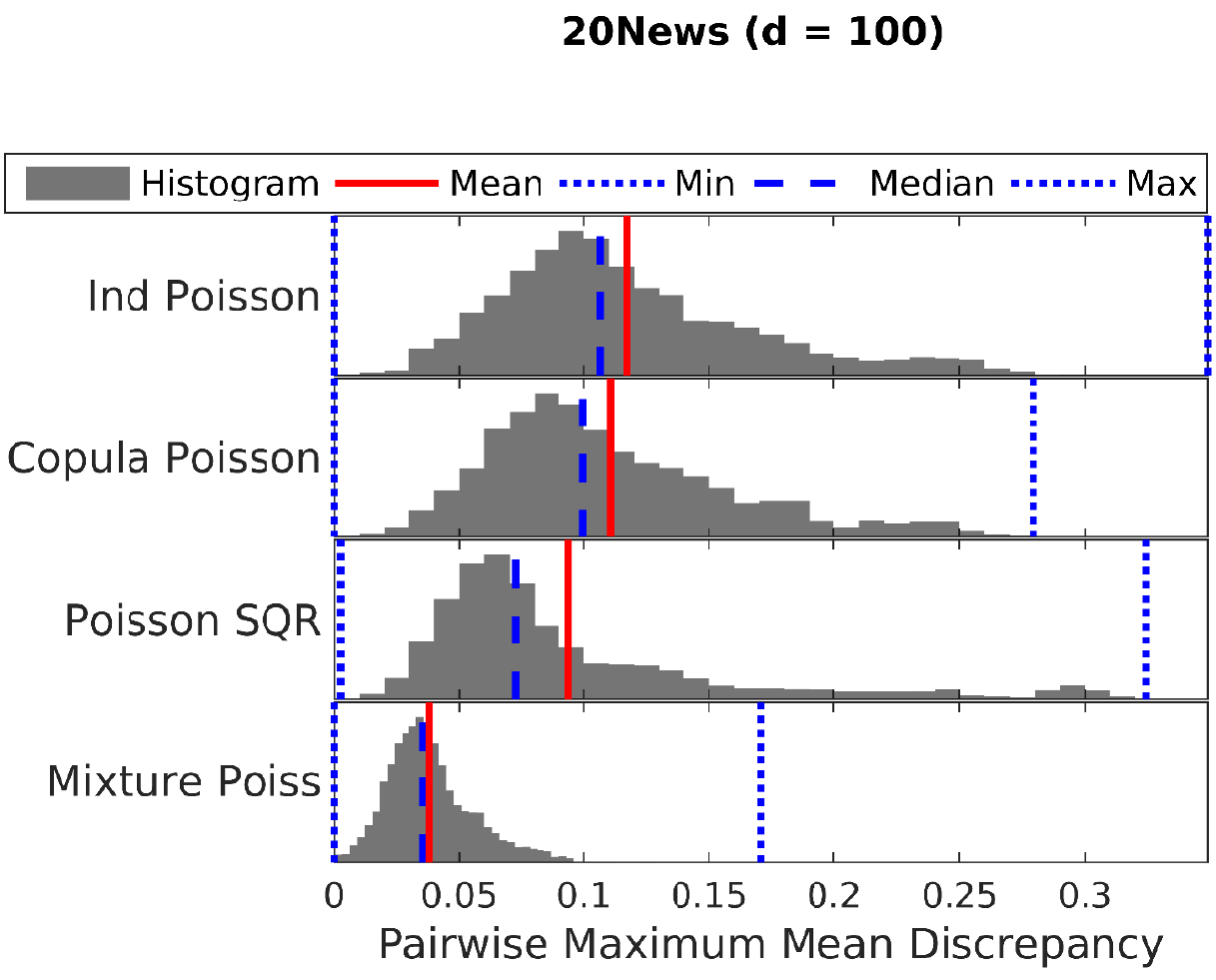} \\
\includegraphics[width=\histwidth]{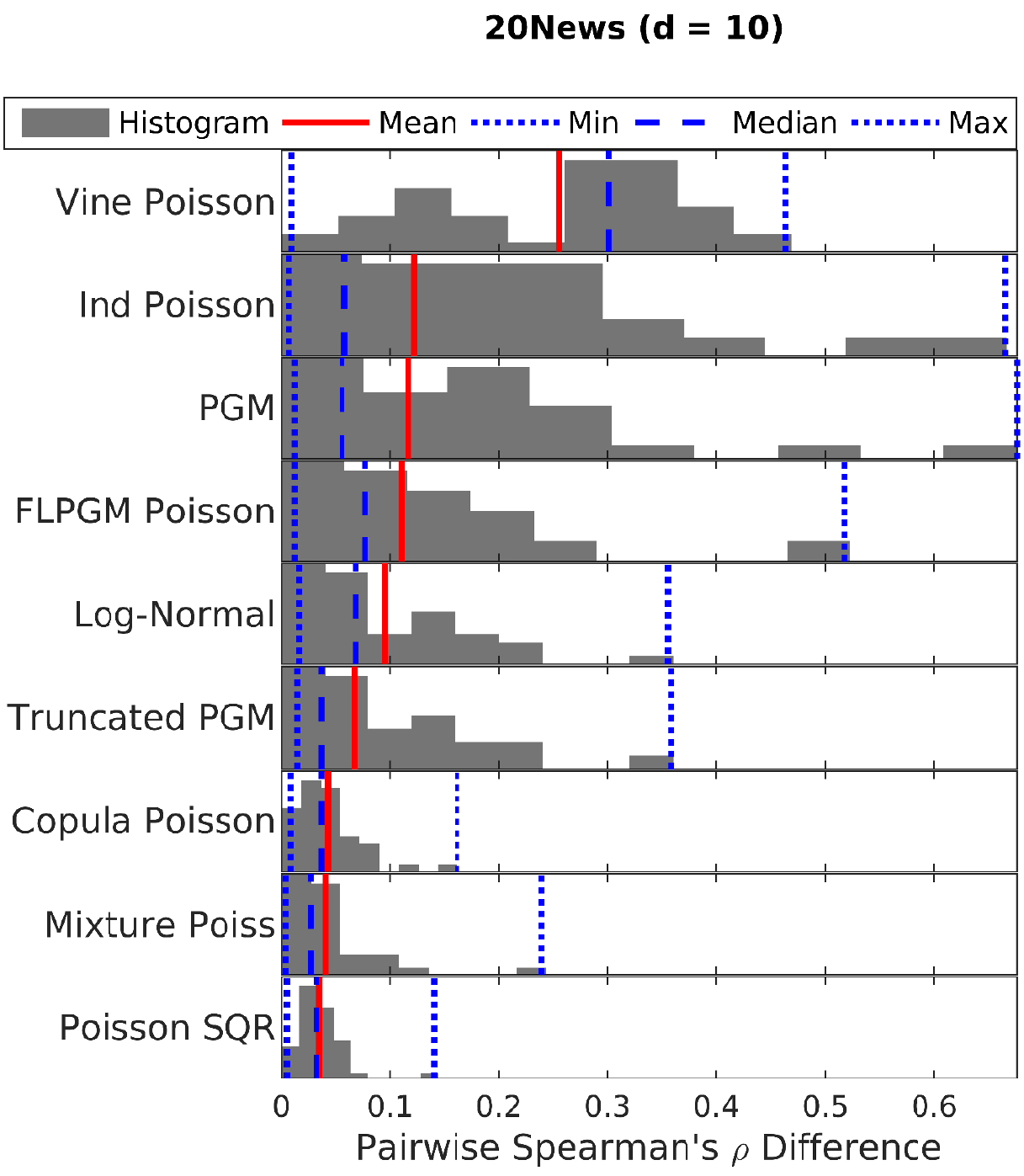}
\includegraphics[width=\histwidth]{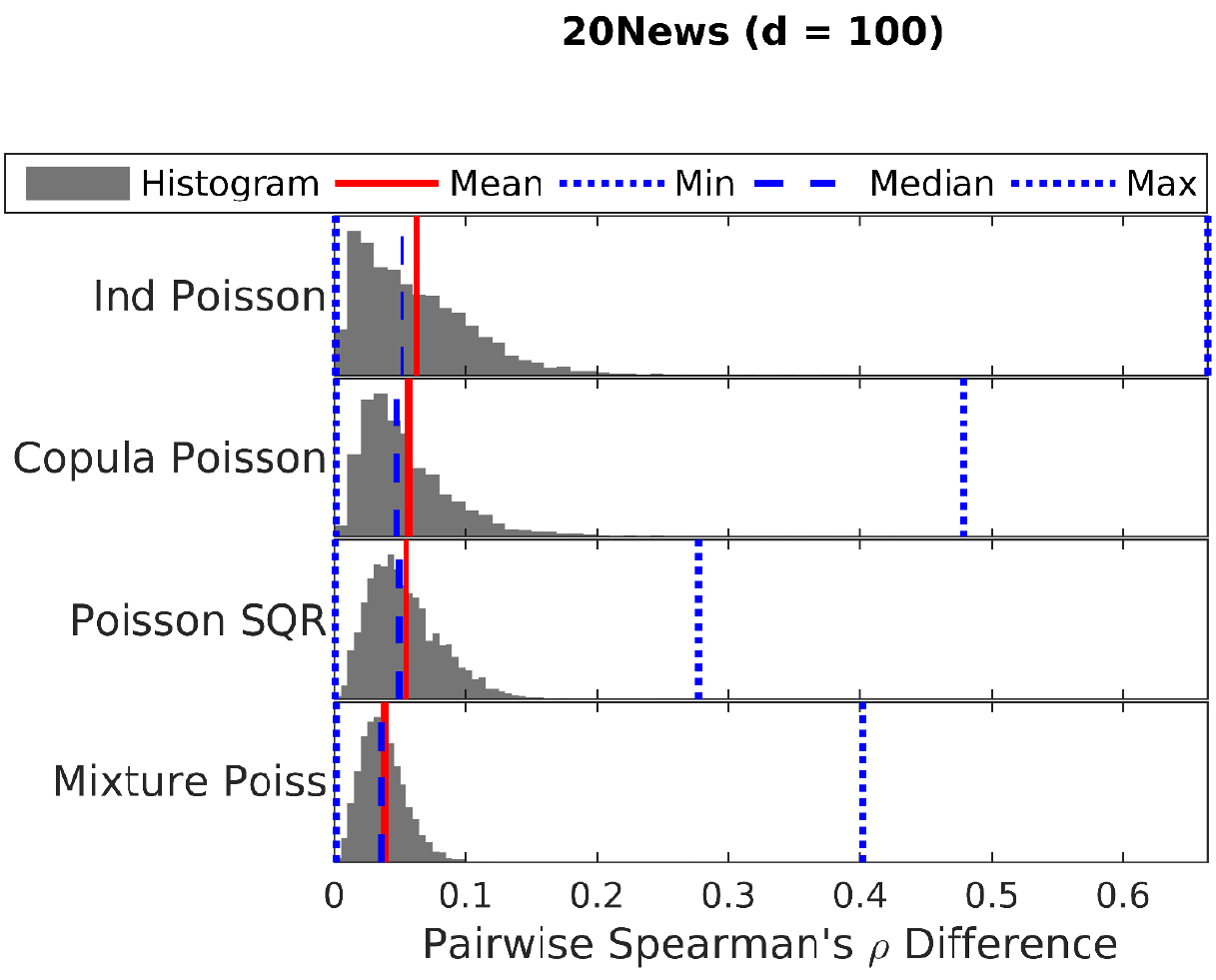}
\caption{The results for the 20 Newsgroup dataset with low count values and medium overdispersion behave very similarly to the results from the Classic3 dataset described in the paper.}
\label{fig:20news}
\end{figure}

We describe and give results for a crime statistics dataset and the 20 Newsgroup dataset.  As mentioned in the paper, these datasets behave similarly to the the BRCA and Classic3 datasets respectively but we include them here for completeness and for additional evidence of the observations described in the paper.  The dataset statistics can be seen in Table~\ref{tab:supp-datasets}.  The results for the crime statistics can be seen in Fig.~\ref{fig:crime-lapd} and the 20 Newsgroup results can be seen in Fig.~\ref{fig:20news}.

\begin{enumerate}
 \item {\bf Crime count dataset (Medium counts, medium overdispersion):} Aggregated crime counts from LAPD during the years 2012-2015.\footnote{\url{https://data.lacity.org/A-Safe-City/Crimes-2012-2015/s9rj-h3s6}. We removed year 2013 and November of 2015 which both clearly had a different distribution than other years likely due to different classification systems.}  The original dataset contains 151 types of crime counts such as ``Burglary'' and ``Vandalism''. This dataset exhibits a wide range of mean values with weak correlation and weak overdispersion.
\item  {\bf 20 Newsgroup text dataset (Low counts, medium overdispersion):} Standard text corpus for document classification with almost 1000 forum posts from 20 different newsgroups. \footnote{\url{http://qwone.com/~jason/20Newsgroups/} We slightly modified this dataset by removing words that were merely for structure or were clearly outliers: ``line'', ``subject'', ``organ'', ``re'', ``post'', ``host'', ``nntp'', and ``maxaxaxaxaxaxaxaxaxaxaxaxaxaxax''.  The raw dataset contained very strong outliers, e.g. the word ``1'' had a mean of 0.58 and a standard deviation of 5.89 but had a maximum value of 344---more than 50 standard deviations away from the mean.  Thus, for each variable, we truncated the values beyond the the 99.5th percentile to be the 99.5th percentile; thus, at most 0.5\% of values were truncated per variable. }
\end{enumerate}

\begin{table}[!ht]
\centering
\caption{Dataset Statistics}\label{tab:supp-datasets}
\vspace{0.5em}
\resizebox{1\textwidth}{!}{
\begin{tabular}{llllllllllllll}
\toprule
\multicolumn{3}{r}{(Per Variable $\Rightarrow$)} & \multicolumn{3}{c}{Means} & \multicolumn{3}{c}{Dispersion Indices} & 
  \multicolumn{3}{c}{Spearman's $\rho$} \\
\cmidrule(r){4-6} \cmidrule(r){7-9} \cmidrule(r){10-12}
Dataset & $\nvar$ & $\ninst$ & Min & Med & Max & Min & Med & Max & Min & Med & Max \\
\midrule
Crime LAPD & 10 & 1035 & 25 & 39 & 118  & 1.4 & 2.1 & 6.2  & -0.19 & 0.14 & 0.52  \\
 & 100 & 1035 & 0.06 & 0.77 & 118  & 0.91 & 1.3 & 16  & -0.49 & 0.02 & 0.78  \\
20News & 10 & 18846 & 0.36 & 0.5 & 1.4  & 0.59 & 1.7 & 6.2  & -0.37 & 0.03 & 0.67  \\
 & 100 & 18846 & 0.07 & 0.15 & 1.4  & 0.83 & 1.9 & 6.2  & -0.37 & 0.05 & 0.67  \\
\bottomrule
\end{tabular}
}
\end{table}

\section{Implementation Details}
\subsection{Copulas Paired with Poison Marginals}
As stated in the paper, we estimated the copula-based models using the Inference Function for Margins (IFM) method via the distributional transform.  More specifically, we first estimated the Poisson marginal distributions. Then, we computed the distributional transform  to map the data from the discrete domain to the continuous domain, i.e. $u = (F(x) + F(x-1))/2$ where $F(\cdot)$ is the Poisson CDF.\footnote{\add{We chose the DT transform because it is computationally and conceptually the simplest of estimation methods even though more complex methods exist for a small number of samples \citep{Nikoloulopoulos2016}.}}  Finally, we estimate the copula distribution using either the \texttt{copulafit} function in MATLAB or the \texttt{RVineStructureSelect} function in the \texttt{VineCopula}\footnote{\url{https://cran.r-project.org/web/packages/VineCopula/index.html}} R package for the Gaussian and vine copulas respectively.  For the vine copula, the vine structure and bivariate copulas were automatically selected in the \texttt{RVineStructureSelect} function; we allowed the following six bivariate copulas and their rotations: Gaussian copula, Student's $t$ copula, Clayton copula, Gumbel copula, Frank copula, and Joe copula.

\subsection{Mixture Models}
For the finite mixture of independent Poissons, we initialized the EM algorithm with the best of 10 $k$-means clusterings and set the maximum number of EM iterations to one hundred. Note that even in high dimensions, the EM algorithm usually converged in under 20 iterations.  For the log-normal mixture, we used 1000 iterations and 400 burn-in iterations for the MCMC algorithm.

\subsection{Conditional Models}
We set the truncation value $R$ to the 99th percentile of the non-zeros in the training dataset. This helped avoid expensive computations if there was one or two very large outliers since each gradient iteration requires $R$ exponential evaluations per observation. We tested 10 regularization parameters of log-spaced points between $\lambda_{\text{max}}$ and $0.0001\lambda_{\text{max}}$ where $\lambda_{\text{max}}$ is the max value of the off diagonals of the training data empirical second moment matrix. In the case of the Poisson SQR, we set $\lambda_{\text{max}}^\text{SQR} = \sqrt{\lambda_{\text{max}}}$ because the sufficient statistics are square roots of the original sufficient statistics.  Essentially, this initially estimates an independent model and slowly moves toward a highly dependent model by reducing the regularization parameter.

\section{Sampling Details}
For the \add{models based on pairing copulas with Poisson marginals,} \remove{copula models,} we \remove{merely} \add{first} sampled from the copula either using the \texttt{copularnd} MATLAB function in the case of the Gaussian copula or \texttt{RVineSim} from the VineCopula R package in the case of the vine copulas; \add{then, we transformed the copula samples to the discrete domain using the Poisson marginal CDFs.} For the mixture models, sampling is also straightforward; we sampled the Poisson mean from the finite mixture or log-normal distribution and then sampled a Poisson variable given this mean.  For the PGM, TPGM and Poisson SQR models, we used Gibbs sampling with 5,000 iterations.  Because the Poisson SQR conditionals are non-standard, we implemented the Gibbs iterations using two steps of Metropolis-Hastings rejection sampling.  For the FLPGM models, we used the annealed importance sampling routines provided by the authors of \citep{IRD15} with 100 annealing steps.  Overall, the copula\add{-based} and mixture models have direct sampling routines whereas the conditional models have natural procedures for Gibbs sampling.

\end{document}